\documentclass[aps,
superscriptaddress,floatfix,12pt]{revtex4}

\usepackage[dvips]{graphicx}
\usepackage{amsmath,amsfonts,bm}
\usepackage{psfig}

\begin{document}

\newcommand\<{\langle}
\renewcommand\>{\rangle}
\renewcommand\d{\partial}
\newcommand\LambdaQCD{\Lambda_{\textrm{QCD}}}
\newcommand\tr{\mathrm{Tr}\,}
\newcommand\+{\dagger}
\newcommand\Fbar{\overline{F}}
\newcommand\g{g_5}


\newcommand{\drawsquare}[2]{\hbox{%
\rule{#2pt}{#1pt}\hskip-#2pt
\rule{#1pt}{#2pt}\hskip-#1pt
\rule[#1pt]{#1pt}{#2pt}}\rule[#1pt]{#2pt}{#2pt}\hskip-#2pt
\rule{#2pt}{#1pt}}

\newcommand{\Yfund}{\raisebox{-.5pt}{\drawsquare{6.5}{0.4}}}
\newcommand{\Ysymm}{\raisebox{-.5pt}{\drawsquare{6.5}{0.4}}\hskip-0.4pt%
        \raisebox{-.5pt}{\drawsquare{6.5}{0.4}}}
\newcommand{\Ythrees}{\raisebox{-.5pt}{\drawsquare{6.5}{0.4}}\hskip-0.4pt%
          \raisebox{-.5pt}{\drawsquare{6.5}{0.4}}\hskip-0.4pt%
          \raisebox{-.5pt}{\drawsquare{6.5}{0.4}}}
\newcommand{\Yfours}{\raisebox{-.5pt}{\drawsquare{6.5}{0.4}}\hskip-0.4pt%
          \raisebox{-.5pt}{\drawsquare{6.5}{0.4}}\hskip-0.4pt%
          \raisebox{-.5pt}{\drawsquare{6.5}{0.4}}\hskip-0.4pt%
          \raisebox{-.5pt}{\drawsquare{6.5}{0.4}}}
\newcommand{\Yasymm}{\raisebox{-3.5pt}{\drawsquare{6.5}{0.4}}\hskip-6.9pt%
        \raisebox{3pt}{\drawsquare{6.5}{0.4}}}
\newcommand{\Ythreea}{\raisebox{-3.5pt}{\drawsquare{6.5}{0.4}}\hskip-6.9pt%
        \raisebox{3pt}{\drawsquare{6.5}{0.4}}\hskip-6.9pt
        \raisebox{9.5pt}{\drawsquare{6.5}{0.4}}}
\newcommand{\Yfoura}{\raisebox{-3.5pt}{\drawsquare{6.5}{0.4}}\hskip-6.9pt%
        \raisebox{3pt}{\drawsquare{6.5}{0.4}}\hskip-6.9pt
        \raisebox{9.5pt}{\drawsquare{6.5}{0.4}}\hskip-6.9pt
        \raisebox{16pt}{\drawsquare{6.5}{0.4}}}
\newcommand{\Yadjoint}{\raisebox{-3.5pt}{\drawsquare{6.5}{0.4}}\hskip-6.9pt%
        \raisebox{3pt}{\drawsquare{6.5}{0.4}}\hskip-0.4pt
        \raisebox{3pt}{\drawsquare{6.5}{0.4}}}
\newcommand{\Ysquare}{\raisebox{-3.5pt}{\drawsquare{6.5}{0.4}}\hskip-0.4pt%
        \raisebox{-3.5pt}{\drawsquare{6.5}{0.4}}\hskip-13.4pt%
        \raisebox{3pt}{\drawsquare{6.5}{0.4}}\hskip-0.4pt%
        \raisebox{3pt}{\drawsquare{6.5}{0.4}}}
\newcommand{\Yflavor}{\Yfund + \overline{\Yfund}} 
\newcommand{\Yoneoone}{\raisebox{-3.5pt}{\drawsquare{6.5}{0.4}}\hskip-6.9pt%
        \raisebox{3pt}{\drawsquare{6.5}{0.4}}\hskip-6.9pt%
        \raisebox{9.5pt}{\drawsquare{6.5}{0.4}}\hskip-0.4pt%
        \raisebox{9.5pt}{\drawsquare{6.5}{0.4}}}%

\preprint{WM-05-???}

\affiliation{Department of Physics, College of William and Mary,
Williamsburg, Virginia 23187-8795}
\title{SUSY Moose Runs and Hops:\\ 
An extra dimension from a broken deformed CFT}
\author{Joshua Erlich and Jong Anly Tan}
\affiliation{Department of Physics, College of William and Mary,
Williamsburg, Virginia 23187-8795}
\email{erlich@physics.wm.edu,jmtanx@wm.edu}

\newcommand\sect[1]{\vspace{\baselineskip}\noindent{\bf{#1}}\vspace{
\baselineskip}}

\begin{abstract}
We find a class of four dimensional deformed conformal field theories which 
appear
extra dimensional when their gauge symmetries are
spontaneously broken.  The theories are supersymmetric moose models which 
flow to interacting conformal fixed points at low energies, deformed by
superpotentials.  Using a-maximization
we give strong nonperturbative evidence that the hopping
terms in the resulting latticized action are relevant deformations of the
fixed point theories.  
These theories have an
intricate structure of RG flows between conformal fixed points.  
Our results suggest that at the stable fixed points 
 each of the bulk gauge couplings and superpotential hopping terms
is turned on, in favor of the extra dimensional interpretation of the
theory.  However, we argue that the higher dimensional gauge coupling is
generically small compared to the size of the extra dimension.
In the presence of a brane the topology of the extra dimension is determined 
dynamically and depends on the
numbers of colors and bulk and brane flavors, 
which suggests phenomenological applications.  
The RG flows between fixed points
in these theories provide a class of tests of Cardy's conjectured a-theorem.
\end{abstract}
\maketitle

\section{Introduction}%

The 
possibility of extra spatial dimensions beyond the three which we
commonly observe has provided a provocative paradigm for model building in
particle physics and gravity in recent years \cite{ADD,AADD,RS}.  
However, due to the non-renormalizability of
higher-dimensional gauge theories they are necessarily low energy
effective theories.  An
ultraviolet completion of higher-dimensional theories was provided by the
deconstruction approach \cite{decon1,decon2}.  In a deconstructed extra
dimension, the extra dimension appears as a lattice of
(3+1)-dimensional gauge theories connected to one another through
link fields transforming in the bifundamental representation of the 
gauge groups at neighboring lattice sites.  For
non-supersymmetric gauge theories the UV completion may be provided by an
asymptotically free gauge theory in which fermions condense to provide the
link fields as nonlinear sigma-model fields.  Alternatively, the link fields 
can arise from fundamental 
scalar fields which obtain vacuum expectation values 
leaving an effective theory which appears higher dimensional below some scale.  
However, scalar fields suffer the usual
problem that their expectation values are naturally of order the cutoff scale,
and supersymmetry (SUSY) is therefore favored to restore naturalness
in this linear sigma model approach.

In this paper we study a class of deconstructed five-dimensional (5D)
supersymmetric gauge theories.
The latticized higher-dimensional theories contain hopping terms in the
Lagrangian which are the latticized kinetic terms containing derivatives in
the extra dimensions.  The scalar link fields obtain vacuum expectation values
(VEVs) 
which break the product gauge symmetry partly or completely.  The spectrum
of gauge bosons mimics the Kaluza-Klein (KK)
spectrum of the extra dimension up to
a scale which depends on the size of the extra dimension $R$ and the 
lattice spacing $l$.  The scale above which the spectrum ceases to be 
higher-dimensional is roughly $\Lambda_{KK}\equiv R^2/l$ for a flat extra
dimension.  Above that scale the
theory is a 4D gauge theory with $k\sim R/l$ factors of the 5D gauge
group.  There are two other
dimensionful scales to keep in mind: the dynamical strong coupling scale of
each 4D gauge group factor $\Lambda_{QCD}$, which
we assume for simplicity
to be the same for each of the gauge groups in the deconstructed 
theory; and the ultimate cutoff of the deconstructed theory $\Lambda_4$, which
must be much higher than $\Lambda_{KK}$ and $\Lambda_{QCD}$
for the deconstructed theory
to provide a UV completion of the 5D theory.  

\begin{figure}
\includegraphics[scale=1]{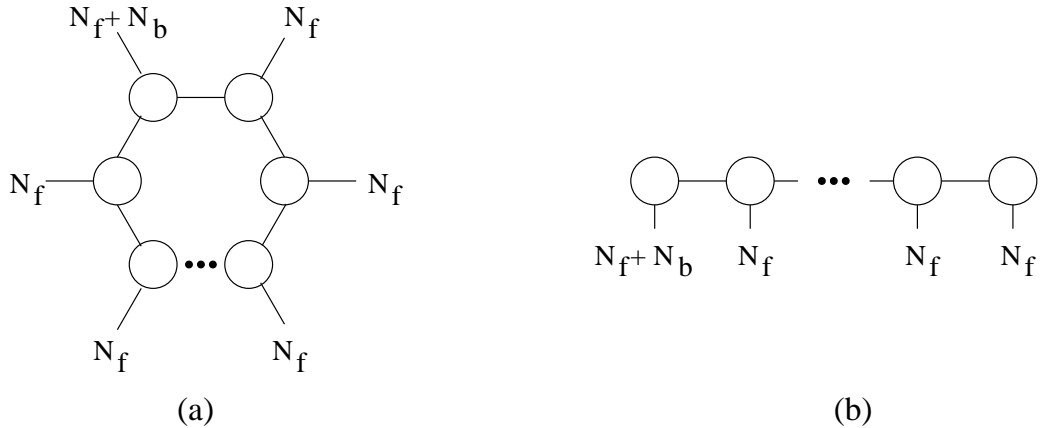}
\caption{\label{fig:moose} a) Circular moose with brane. b) Interval moose with
brane.}
\end{figure}

Flavors in the
bulk of the extra dimension are added to the deconstructed theory by 
replicating the flavors at each lattice site and including a hopping
superpotential \cite{SUSY-dec}.  Flavors on a 3-brane are modeled by 
localizing those flavors to a single lattice site.  The moose (or quiver)
diagram representing the deconstructed theory on a circle with $N_f$ bulk 
flavors and $N_b$ flavors localized to a 3-brane
is given in FIG.~\ref{fig:moose}a.  The moose diagram for a 
theory on an interval with extra 
flavors on one boundary is shown in FIG.~\ref{fig:moose}b.
Nodes in the moose diagram correspond to SU$(N)$ gauge groups, and lines
correspond to fields transforming in the fundamental representation of the
gauge groups at the nodes on which those lines end.  We do not distinguish
fundamental and antifundamental representations in the moose diagrams as it
is conventional to assign links between nodes the bifundamental representation
$(\Yfund,\overline{\Yfund})$, and the charges are otherwise determined by
gauge anomalies.

The hopping terms in the action for bulk flavors are reproduced by a 
superpotential, as we review below.  In the absence of the superpotential
the models that we study have a large number of interacting
conformal fixed points. At each
fixed point some combination
of the independent gauge couplings is nonvanishing.  We use the 
a-maximization procedure of Intriligator and Wecht
\cite{a-max} to calculate the dimensions of operators at these fixed points.
Depending on the numbers of colors and bulk and brane flavors 
we determine which
of the conformal fixed points are stable to gauging of the ungauged symmetry 
groups at those fixed points, and we argue that generically  all 
of the bulk gauge couplings are 
nonvanishing at the stable 
fixed points.  

We will show that for the braneless moose
the hopping superpotential is composed of relevant
operators in the neighborhood of the fixed points, so that those 
operators will generically have 
nonvanishing couplings in the infrared as required for the extra dimensional
interpretation of the moose model near its stable fixed point.  Furthermore, we
will discover that the topology of the deconstructed extra dimension depends
on the numbers of colors and bulk and brane flavors.

\subsection{The Model}

For definiteness we will mostly consider 
the circular moose model with $N_f$ bulk flavors and $N_b$ brane 
flavors, as in FIG.~\ref{fig:moose}a.  The gauge groups and matter content
are summarized in the table below:
\begin{eqnarray}
&&\begin{array}{c|ccccc|c}
  & SU(N)_1 & SU(N)_2 & SU(N)_3 & \cdots & SU(N)_k & U(1)_R \\
\hline
  Q_1 & \Yfund  & \overline{\Yfund}  & 1       & \cdots & 1 & r_{Q_1} \\
  Q_2 & 1       & \Yfund  & \overline{\Yfund}  & \cdots & 1 & r_{Q_2}\\
  \vdots & \vdots & \vdots & \vdots & \ddots & \vdots & \vdots\\
  Q_k & \overline{\Yfund} & 1 & 1 & \cdots & \Yfund & r_{Q_k}\\ \hline 
(N_f+N_b)\ F_1 & \Yfund & 1 & 1 & \cdots & 1 & r_{F_1}\\
(N_f+N_b)  \overline{F}_1 & \overline{\Yfund} & 1 & 1 & \cdots & 1 & r_{\overline{F}_1}\\
N_f\  F_2 & 1 & \Yfund & 1 & \cdots & 1 & r_{F_2}\\
N_f\  \overline{F}_2 & 1 & \overline{\Yfund} & 1 & \cdots & 1 & r_{\overline{F}_2}\\
   \vdots & \vdots & \vdots & \vdots & \ddots & \vdots & \vdots\\
N_f\  F_k & 1 & 1 & 1 & \cdots & \Yfund & r_{F_k}\\
N_f\  \overline{F}_k & 1 & 1 & 1 & \cdots & \overline{\Yfund} & r_{\overline{F}_k}\\ 
\hline
\end{array} \label{moosetable} \end{eqnarray}
The $Q_i$ are the link fields, and $F_i+\overline{F}_i$ are the flavors.  They
are each chiral multiplets of 4D ${\cal N}$=1 SUSY.  The brane is on the first
lattice site, corresponding to the gauge group factor SU$(N)_1$.

The moduli space of this theory allows the link fields to obtain expectation
values, which we assume are at least approximately proportional to
the identity in the SU$(N)\times$SU$(N)$ gauge group space under which each
link field transforms: $\left<Q_i
\right>=v_i \boldsymbol{1}$.  The gauge group is then broken to the
diagonal SU$(N)$ subgroup at low energies.
If there is a superpotential in this model of the form,
\begin{equation}
W_{\rm flavor}= \sqrt{2} g \sum_i
{\rm tr} (\overline{F}_i Q_i F_{i+1}) + \sum_i m_i F_i \overline{F}_i,
\label{eq:superpot}\end{equation}
then the theory has an accidentally enhanced ${\cal N}$=2 SUSY in the
IR.  This is the most general perturbatively renormalizable
superpotential that one can add to the theory, except that the
coefficient of the cubic term depends on the gauge coupling $g$ in
order to have the enhanced SUSY required for minimal 5D SUSY in the
effective theory.

The hopping terms for the link fields
$Q_i$ arise from the D-terms when the scalar component of the link 
fields obtain
VEVs, $\left<Q_i\right>\approx v_i\boldsymbol{1}$
(\cite{SUSY-dec}). 
In general, it
may not be possible to choose the VEVs exactly proportional to the identity
because such vacua may not belong to the moduli space of the theory.  
In such cases, the
deviation from the identity is a 
reflection of the low energy dynamics below
the scale $1/R$ \cite{SUSY-instantons}.
If all of the
vevs are equal (and nonvanishing) 
the hopping terms include the latticized kinetic terms of a
flat extra dimension.  
The flavor
mass terms are also tuned so that $m_i=-\sqrt{2} g \left<Q_i\right>$  in
(\ref{eq:superpot}).  We will have little new to say about this fine tuning.
If the VEVs differ in a smooth fashion from one site
to the next then the hopping terms produce a warped extra dimension 
\cite{warped-dec}.  Warped deconstruction models can do interesting things,
including breaking of supergravity without breaking global SUSY \cite{brian}.

The VEVs are not determined {\em a priori} as there is a continuous moduli 
space.  
In the absence of flavors the circular moose theory has a Coulomb branch on
which the gauge symmetry is broken to U(1)$^{N-1}$ \cite{CEFS}.  
For a flat extra dimension
the continuum limit
requires taking the number of lattice sites $k$ to be large, with the size
of the extra dimension $R=k/(gv)$  and the 5D gauge coupling $\g^2=g^2/(kR)$
fixed with respect to the dynamical scale of the low energy diagonal SU($N$)
gauge theory.  
This requires $v\sim\sqrt{k}$ and $g\sim\sqrt{k}$ as $k$ becomes 
large.
Including flavors, Higgs branches open up along which the flavors obtain
VEVs.  If the flavor VEVs are
small enough, they may be considered a reflection of the dynamics below the
scale $1/R$.  The portions of the Higgs branches with larger flavor VEVs
(compared to $\Lambda_{KK}$), though present, do not correspond to
an extra dimension so we will not discuss them here.  

\subsection{Deconstruction as a UV Completion}

Models of deconstructed extra dimensions are useful for two reasons:
they provide a definition of 5D gauge theories which are otherwise not
renormalizable; and they allow for 4D model building while maintaining
the desirable features of an extra dimension.  In this paper we focus
on the former aspect, renormalizability of the deconstructed theory, although
we also discuss the novel phenomenology of our deformed conformal moose models.
The
hopping potential which provides the latticized kinetic terms
is composed of marginal and relevant operators at
tree level, and
it is possible {\em a priori} that some of the terms in 
the potential become irrelevant at low energies due
to quantum fluctuations.  Whether or not this disastrous possibility
(for the extra dimensional interpretation) 
occurs can be addressed perturbatively
if the coupling is weak.  At strong coupling, however, nonperturbative methods
are necessary to study the theory.  For supersymmetric theories such 
nonperturbative methods exist \cite{SUSY-review}
and have already proven useful for studying
certain deconstructed gauge theories.  In particular, 
4D instanton effects in the deconstructed 5D ${\cal N}$=1 supersymmetric
5D pure gauge theory were examined in \cite{SUSY-instantons}.  
The 6D theory
with (0,2) supersymmetry
and the 6D little string theory were studied in \cite{0-2}, and the 
Seiberg-Witten approach gave new nonperturbative insights in \cite{0-2-SW}.

Strictly speaking, the techniques we use only 
apply to the origin of moduli space, with vanishing VEVs.  Hence, we are
only considering energies $> \Lambda_{KK}$ so that the VEVs are
negligible near the (approximately) fixed points.  In the continuum limit
we need to take the couplings of individual gauge group
factors to be large.  Since
$\Lambda_{QCD}$ is the strong coupling scale of each 4D
gauge group factor, we assume $\Lambda_{QCD}>\Lambda_{KK}$.  
Hence, we expect the 
renormalization group flows to generically approach the
approximate fixed 
point at energies higher than $\Lambda_{KK}$, and our analysis
remains valid.  It is also
interesting to consider the evolution of operators when flowing to
still smaller scales.  This evolution 
requires matching of operators while the heavier
KK modes are integrated out, and we will not address the issue here.

Although the superpotential is not renormalized, wavefunction
renormalization causes the flavor hopping terms
to run with energy scale.
It is in general not possible to calculate the anomalous dimensions of the 
hopping terms in the Lagrangian at strong coupling.  However, the 
conformal anomaly and the R-current anomaly are in a supersymmetry multiplet, 
with the consequence that the conformal dimension of an operator is 
proportional to its R-charge.
At a conformal fixed point, the R-charges are in turn related
to central charges in the theory, which can sometimes be determined exactly
\cite{Freedman}.  
The problem is that quite often there is not a unique gauge
anomaly free R-symmetry,
so the goal is then to determine which of the continuum of
anomaly free R-symmetries is related
by supersymmetry to the conformal anomaly.  Without repeating the clever
analysis of
\cite{a-max}, the remarkably simple procedure that determines the correct
R-charges for this purpose is to maximize the Euler anomaly, \begin{equation}
a=\frac{3}{32}\tr\left( 3 U(1)_R^3-U(1)_R\right), \end{equation}
where the trace is over the fermions in chiral or vector multiplets, and 
the maximization is over all gauge anomaly free R-symmetries. This procedure
is known as a-maximization. 
The procedure may fail if accidental symmetries
appear at low energies, unless all such accidental symmetries have been 
identified and accounted for.  Indeed, there are many examples of 
supersymmetric gauge theories for which the existence of accidental symmetries
is required for consistency of Seiberg duality and unitarity \cite{accidental},
and we will have to account for such accidental symmetries in our moose models.

\subsection{Fixed Point Stability}

Models with product gauge groups and bifundamental fields
have been studied nonperturbatively using 
Seiberg duality techniques \cite{Poppitz}, Seiberg-Witten theory \cite{CEFS},
and a-maximization \cite{Barnes,deconfined}.  
The conclusion of these analyses is that
these 
theories
have an intricate vacuum structure which
depends qualitatively on the numbers of flavors and colors of each gauge group
factor.  In \cite{Poppitz} it was concluded that theories with two gauge groups
can have a variety of conformal fixed points with either or both of the
gauge couplings nonvanishing.  IR stability of these 
fixed points depends
on the numbers of flavors and colors.  If the conformal
dimensions of the fields are
known at a fixed point for which at least one of the gauge couplings vanishes, 
then the NSVZ beta-functions for the remaining gauge couplings 
are also known in a neighborhood of that fixed
point (perturbatively in the small 
coupling(s)).  The signs of the beta-functions then determine stability of the 
fixed point to perturbations of the gauge couplings.  One can thereby 
determine which of the fixed points a generic flow from the UV will approach
in the IR.  Superpotential deformations of the theory can be studied
similarly
to see whether the operators appearing in the deconstructed superpotential
are relevant perturbations of the theory in their absence, as is necessary for
the theory to appear higher dimensional below the scale $\Lambda_{KK}$.  

At fixed 
points with more than one nonvanishing gauge coupling
a-maximization is  generally required to determine operator dimensions.
This analysis was done for a class of models with two 
gauge groups in \cite{Barnes}.  We will extend the analysis to models with
more gauge group factors, and focus our study to models related to 
deconstructed extra dimensions.  As described earlier,
we assume that each of the
gauge group factors has the same number of colors $N$, and all but one
of the gauge group factors has the same number of flavors $N_f$.  At one
special lattice site, which we label the first lattice site and refer to as 
the brane, we allow for an
additional $N_b$ flavors representing brane-localized matter.

We will argue that generically the bulk gauge couplings are
nonvanishing at the stable fixed points.  In the absence of a brane, 
at perturbative stable fixed points
near the edge of the conformal window we will find that the gauge couplings
are all equal to one another.  In the
presence of a brane we will not determine the gauge couplings, but we
will find that depending on the numbers of colors, bulk and brane
flavors, the gauge coupling at the brane may vanish at the stable
fixed point.  We will conclude from this that the topology of the
deconstructed extra dimension depends on the numbers of flavors and
colors.

A study of the  flows between conformal fixed points in 
these theories allows for a large class of nontrivial tests 
of the ``a-theorem.''
The a-theorem is a 
conjecture by Cardy that the Euler anomaly $a$ plays
the role of a Zamolodchikov
$C$-function in four dimensions \cite{Cardy}.  Heuristically, the $C$-function
effectively counts the number of degrees of freedom in the 
theory as a function of energy scale, 
which is expected to
decrease as degrees of freedom are integrated out.  As such, the Euler anomaly
is supposed to be always positive and
flow from higher values to smaller values as energy is decreased.  Neither of
these requirements ({\em i.e.} positivity or monotonicity) is known to be
guaranteed, although there have been partially successful attempts at proving
them, 
and many tests have been provided \cite{Freedman,Barnes,athm-tests}.  We
will test Cardy's conjecture in a large class of flows between fixed points, 
and will find consistency with the conjecture in each case.  We will find that
the a-theorem also precludes certain types of RG flows in these models that are
otherwise difficult to argue against.

\section{The Braneless Moose}
In this section we consider a deconstructed extra dimension with SU($N$) gauge
group and $N_f$ bulk flavors, but no localized flavors on a brane,
$N_b=0$.  For now we turn off the superpotential, so
that the flavors do not yet hop in the extra dimension.  Later we will check
that the superpotential in the deconstructed theory is a relevant deformation
of the theory.  In the
absence of a brane, the $\mathbb{Z}_k$ symmetry of lattice translations
puts all of the bifundamental fields $Q_i$ on equal footing, and similarly for
the flavors ($F_i+\overline{F}_i$). If the $\mathbb{Z}_k$ symmetry is preserved
at a fixed point, then all of the
gauge couplings at the fixed point will be
equal, $g^*_i\equiv g^*$, 
the R-charges of the bifundamental fields will be identical to one another, 
and the R-charges
of the flavors will similarly be identical to one another.  
At such a fixed point, the 
theory is as in the table (\ref{moosetable}), with $N_b=0$, 
$r_{Q_i}\equiv r_Q$, and $r_{F_i}=r_{\overline{F}_i}\equiv r_F$.
We require
that $N/2<N_f<2N$ so that each SU$(N)$ is in its naive conformal window.  
We call it the naive conformal window because the actual conformal window for
this class of theories may extend to $N_f<N/2$ because of the additional
gauge interactions \cite{Barnes}.  For simplicity we focus our study on the
naive conformal window, and leave the analysis of the larger conformal window
for future work.
Anomaly cancellation requires (assuming $r_F=r_{\overline{F}}$, which is 
easily checked to be valid here), \begin{equation}
N(r_Q-1)+N_f(r_F-1)+N=0.
\label{eq:anom-nobrane}\end{equation}
The Euler anomaly in this case is given by, \begin{equation}
a=\frac{3}{32}\left(kN^2\left(3(r_Q-1)^3-(r_Q-1)\right)+2kN_f N\left(3(r_F-1)^3-(r_F-1)\right)+
2k(N^2-1)\right).
\end{equation}
Maximizing $a$ 
subject to the constraint (\ref{eq:anom-nobrane}),
we find that the R-charges which belong to the superconformal algebra and
determine operator dimensions are given by, \begin{eqnarray}
r_Q^*&=& \frac{-3 N_f^2+N_f\sqrt{20 N^2 -N_f^2}}{6 N^2-3N_f^2}\nonumber \\
r_F^*&=& \frac{6 N^2+3 N N_f-3 N_f^2-N \sqrt{20 N^2-N_f^2}}{
6 N^2-3 N_f^2} \label{eq:Zk-R}\end{eqnarray}
and the value of $a$ at the fixed point is, \begin{eqnarray}
a^*&=&
-\frac{ k \left(9 N_f^4+N^4 \left(36+90 N_f^2-20 N_f\sqrt{20 N^2 -
N_f^2}\right)\right)}{
48 \left(-2 N^2+N_f^2\right)^2} \nonumber \\
&&-\frac{k N^2 N_f^2 
\left(-36-18 N_f^2+N_f\sqrt{20 N^2 -N_f^2}\right)}{
48 \left(-2 N^2+N_f^2\right)^2}
\end{eqnarray}

Although the denominators in the expressions above vanish at
$N_f=\sqrt{2}N$, $a^*$ is a regular function of $N$ and $N_f$ there.
The values of the fixed-point R-charges were obtained assuming the absence
of an accidental symmetry that could mix with the R-symmetry in the
infrared.  We can check the consistency of this assumption by testing whether
or not any gauge invariant chiral
operator would have dimension $<1$ with these assignments, in
violation of unitarity \cite{unitarity,Seiberg}.  
We find that operators of the form $\Fbar F$ would
fall out of the unitarity bound for $N_f<\frac{1}{2}\left(3N-\sqrt{7}N\right)$
or $N_f>\frac{1}{2}\left(3N+\sqrt{7}N\right)$, which lie outside of the
naive conformal window $N/2<N_f<2N$.  With more than two lattice sites we find
that there is no operator which would violate unitarity with the R-charge
assignments above in the naive conformal window. 
In the case of two lattice sites, the operator ${\cal O}=
Q_1 Q_2$ is gauge invariant, and would lead to violation of unitarity if
$N_f<\sqrt{2/5}N$, which lies within the naive conformal window.  
In order
for this scenario to be consistent it is commonly assumed that an accidental
U(1) symmetry appears at low energies under which only the chiral 
superfield associated with the unitarity violating operator
is charged \cite{Seiberg,accidental}.
The accidental symmetry mixes with the R-symmetry and, consistent with
a revised a-maximization, the R-charge of the offending superfield 
then takes its free value of 2/3.
For this purpose, we maximize the corrected expression for $a$, taking
into account the accidental symmetry: \begin{equation}
a_{\rm new}=a_{\rm old}-\frac{3}{32}\left[\left(3(r_{\cal O}-1)^3-(r
_{\cal O}-1)\right)
+2/9\right],
\end{equation}
where 3/32$\times$2/9 is the contribution to $a$ from a free chiral multiplet
with R-charge 2/3.  The result of $a$-maximization including this correction
is not particularly illuminating, except that we have checked 
the R-charges change 
by not more than a few percent when $N=2$ due to the accidental symmetry,
and change by even smaller amounts for larger $N$.

We are interested in the stability of the fixed point found above in the
sense of renormalization group flow.  To study this, we begin by
noting that in the
naive conformal window $N/2<N_f<2N$, there is necessarily a fixed point 
with only one of the $k$ gauge couplings nonvanishing, as this is SUSY
QCD in its conformal window \cite{Seiberg}.  
At such a fixed point the R-charges
for all of the fields are known, so perturbing about the fixed point
we can calculate 
the $\beta$-functions for the remaining weakly gauged SU$(N)$ couplings
nonperturbatively in the strong couplings.
This allows us to determine whether those fixed points are stable in the IR,
and thereby produces a plausible picture of RG flows in the space of 
couplings and fixed points.  This technique was advocated in 
\cite{Poppitz,Barnes}, the only difference here being that with additional
gauge group factors the space of fixed points is more complicated.  We may
or may not also have stable fixed points with two nonvanishing couplings, or
three nonvanishing couplings, {\em etc}.  
We will label fixed points by the number
of nonvanishing gauge couplings, with an A-type fixed point having a single
nonvanishing gauge coupling, a B-type fixed point having two nonvanishing
gauge couplings, and so on.  

\subsection{A-type fixed points}
At a fixed point with a single nonvanishing coupling, say $g_1$,
the interacting part of the theory looks like SU$(N)$
SUSY QCD with $N+N_f$ flavors
\cite{Seiberg}.
There is a unique
anomaly-free R-charge consistent with the enhanced SU$(N+N_f)$ global
symmetry for which the fields $Q_1$, $Q_N$, $F_1$ 
and $\overline{F}_1$ have R-charge, \begin{equation}
r=N_f/(N+N_f). \label{eq:r} \end{equation}  
Weakly gauging SU$(N)_2$, the perturbed theory is given by the table below (for
$k>2$ lattice sites, which we assume throughout this section):
\begin{eqnarray}
&&\begin{array}{c|cc|c}
  & SU(N)_1 & SU(N)_2 & U(1)_R \\
\hline
  Q_1 & \Yfund  & \overline{\Yfund}  & r \\
N_f\ F_1 & \Yfund & 1  & r\\
(N_f+N)  \overline{F}_1 & \overline{\Yfund} & 1 & r\\
(N_f+N)\  F_2 & 1 & \Yfund  & 2/3\\
N_f\  \overline{F}_2 & 1 & \overline{\Yfund} & 2/3\\
\hline
\end{array} \label{Atable} \end{eqnarray}
The NSVZ $\beta$-function for the perturbed
gauge coupling is given by \cite{NSVZ}, 
\begin{eqnarray}
\beta_2&=&-
\frac{g_2^3}{16\pi^2}\frac{\left(3G-\sum_i\mu_i(1-\gamma_i)\right)}{
1-\frac{g_2^2}{8\pi^2}\,G} \nonumber \\
&=&-
\frac{3g_2^3}{16\pi^2}\frac{\left(G-\sum_i\mu_i(1-r_i)\right)}{
1-\frac{g_2^2}{8\pi^2}\,G} \label{eq:beta2}\\
&=&-
\frac{g_2^3}{32\pi^2}\frac{2N^2+3N N_f-2N_f^2}{
\left(N+N_f\right)\left(1-\frac{g_2^2 N}{8\pi^2}\right)} \nonumber
\end{eqnarray}
where $G$ is the quadratic Casimir of the gauge group
(normalized to $N$ for SU$(N)$);
$\mu_i$ is the Dynkin index of the representation of the gauge group
under which the corresponding chiral multiplet transforms ($\mu$=1/2 for the 
fundamental representation); 
$\gamma_i$ is the anomalous dimension of the $i$th 
chiral multiplet, and $r_i$ is its R-charge.
In the second line of (\ref{eq:beta2})
we used the relation between the anomalous dimension
of a chiral superfield and its R-charge, \begin{equation}
{\rm dim}(\phi)=1+\frac{\gamma_\phi}{2}=\frac{3r_\phi}{2}.
\label{eq:R-anomdim}
\end{equation}
In the last line of (\ref{eq:beta2})
we used the fact that $N$ of the fields charged
under SU$(N)_2$ are also charged under SU$(N)_1$ and have R-charge $r$ near
the fixed point, while
the remaining fields charged under SU$(N)_2$ are free with R-charge 2/3.
The IR fixed point is unstable to perturbations by $g_2$ (or $g_N$) if the 
$\beta$-function is negative.  We find $\beta_2<0$ for $N_f<2N$, so the
A-type fixed point is unstable in the entire naive conformal window.
All gauge couplings other than $g_N$, $g_1$ and $g_2$ have negative 
$\beta$-functions because they are asymptotically free SUSY QCD.  
In summary, in the naive conformal window the one-loop $\beta$-functions 
for all weak couplings near an A-type fixed point are negative, 
so that the A-type fixed points are necessarily
unstable to turning on those gauge couplings.  

\subsection{B-type fixed points}
We have found that the A-type fixed points will tend to flow to fixed
points with additional gauge couplings turned on in the infrared.
Here we are interested in fixed points with two nonvanishing couplings, which
we refer to as B-type fixed points.
There are several distinct possibilities for such fixed points.  If the
nonvanishing couplings correspond to  
nodes in the moose which are not
either neighbors or next-to-nearest neighbors, then the fixed point and
its stability analysis
acts as 
two decoupled A-type fixed points.  As such, we will focus our attention
on B-type fixed points which are not equivalent to decoupled 
A-type fixed points.  We
define B1-type fixed points to be those fixed points with two neighboring
gauge couplings turned on, and B2-type fixed points to be those with 
next-to-nearest neighboring gauge couplings turned on.
The low energy interacting theory for a B1-type fixed point
is specified by the following table:
\begin{eqnarray}
&&\begin{array}{c|cc|c}
  & SU(N)_1 & SU(N)_2 & U(1)_R \\
\hline
  Q_1 & \Yfund  & \overline{\Yfund}  & r_{Q} \\
N_f\ F_1 & \Yfund & 1  & r_{F_1}\\
(N_f+N)  \overline{F}_1 & \overline{\Yfund} & 1 & r_{\Fbar_1}\\
(N_f+N)\  F_2 & 1 & \Yfund  & r_{F_2}\\
N_f\  \overline{F}_2 & 1 & \overline{\Yfund} & r_{\Fbar_2}\\
\hline
\end{array} \label{Btable} \end{eqnarray}
At the fixed point, if there are no accidental symmetries then
by maximizing $a$ we find that, \begin{eqnarray}
r_{F_1}=r_{F_2}=r_{\overline{F}_1}=r_{\overline{F}_2}&=& 
\frac{9 N^2-12 N_f^2-N\sqrt{(73 N^2-4 N N_f-4 N_f^2)}}{
3 (N^2-4 N N_f-4 N_f^2)}\label{eq:rF1}\end{eqnarray} \begin{eqnarray}
r_Q&=& 
\frac{-9 N^2-12 N N_f+(2 N_f+N)\sqrt{(73 N^2-4 N N_f-4 N_f^2)}-12 N_f^2}{
3(N^2-4 N N_f-4 N_f^2)} \label{eq:rQ}
\end{eqnarray}
Except for $r_Q$, 
each field has R-charge $>1/3$ at the fixed point in
the naive conformal window, so that 
there is no gauge invariant $\overline{F}_i F_i$ type 
operator which would violate unitarity 
(by having R-charge $<2/3$) in the
absence of an accidental symmetry.   Numerically, one finds that 
the operator $\det Q$ has R-charge $>2/3$ for any $N$, $N_f$ in the naive 
conformal window.
Hence, it is natural to assume that there
is no accidental symmetry in this case.

Perturbing about the fixed point by weakly gauging SU$(N)_3$, we find that 
the  $\beta$-function for the small gauge coupling takes the form,
\begin{equation}
\beta_3^{(B1)}= 
-\frac{g_3^3}{32\pi^2}
\frac{11 N^3-10 N^2 N_f+8 N_f^3- 12 N N_f^2-N^2\sqrt{73 N^2-4 N N_f-4 N_f^2}}{
(N^2-4 N N_f-4 N_f^2)\left(1-\frac{g_3^2 N}{8\pi^2}\right)}
\label{eq:beta3B1}\end{equation}
One can check from this expression that $\beta_3<0$ in the entire naive
conformal window, so that the B1-type fixed point is unstable in the IR.  The
value of $a$ at the B1-type fixed point will be given more generally later,
in (\ref{eq:aa}).

For a B2-type fixed point, the interacting theory is that of two A-type fixed 
points.  The difference in the stability analysis compared with 
the A-type fixed point
is seen by weakly gauging the SU$(N)$ gauge group in the
``middle'' of the two low energy SU$(N)$ gauge group factors.  For example,
if the SU$(N)_1$ and SU$(N)_3$ gauge couplings are nonvanishing at
the fixed point, then weakly gauging SU$(N)_2$ gives rise to the following
perturbed theory, where $r$ is given by (\ref{eq:r}):
\begin{eqnarray}
&&\begin{array}{c|ccc|c}
  & SU(N)_1 & SU(N)_2 & SU(N)_3 & U(1)_R \\
\hline
  Q_1 & \Yfund  & \overline{\Yfund}  & 1 & r \\
  Q_2 & 1 & \Yfund  & \overline{\Yfund} & r \\
N_f\ F_1 & \Yfund & 1  &1 & r \\
N_f\  \overline{F}_1 & \overline{\Yfund} & 1 &1 & r \\
N_f\  F_2 & 1 & \Yfund  &1 & 2/3 \\
N_f\  \overline{F}_2 & 1 & \overline{\Yfund} &1 & 2/3 \\
N_f\  F_3 & 1 & 1& \Yfund  & r\\
N_f\  \overline{F}_3 & 1 & 1  &\overline{\Yfund} & r \\

\hline
\end{array} \label{B2table} \end{eqnarray}

The $\beta$-function for the middle group is then, \begin{eqnarray}
\beta_2^{(B2)}&=&-
\frac{3g_2^3}{16\pi^2}\frac{\left(N-\frac{2N_f}{2}(1-2/3)-N(1-r)\right)}{
1-\frac{g_2^2 N}{8\pi^2}}  \nonumber \\
&=&-
\frac{g_2^3}{16\pi^2}\frac{\left(2N-N_f\right)N_f}{
(1-\frac{g_2^2 N}{8\pi^2})(N+N_f)}  \label{eq:beta2B2}\end{eqnarray}
It follows that $\beta_2^{(B2)}<0$ for $N_f<2N$, and
$B2$-type fixed points are therefore unstable in the entire conformal 
window.

\subsection{General fixed points}
Due to asymptotic freedom of the theory in the naive conformal
window, fixed points are generically unstable to 
the development of gauge couplings which are not 
connected by link fields to any of the interacting gauge groups at the fixed 
point.  This implies that generically
at least one third of the gauge couplings in the moose will be turned on at
the stable fixed point, as each node that is turned off with at least two
nodes turned off on either side of it is itself 
unstable to turning on.  The stability arguments in the previous sections 
suggest that generically an even 
larger fraction of the gauge couplings are turned
on at the stable fixed point, and indeed we conjecture that all of the
gauge couplings are generically turned on.  In all of the cases which we
will study numerically we will find that the bulk gauge couplings are all
turned on, with or without a brane.
For a generic fixed point with a large number of interacting gauge
groups the analysis is complicated.  However, there are a number of generic
statements which can be made.

Suppose the fixed point contains a chain of interacting gauge groups
as in 
FIG.~\ref{fig:InteractingMoose}, where solid nodes correspond to the
interacting
gauge groups at the fixed point, and unfilled nodes are noninteracting at
the fixed point.  Such a fixed point structure is analogous to the B1 type 
fixed point, so we refer to such a fixed point generically as Type 1.
We are interested in the effect of weakly
gauging the group corresponding to the lightly filled 
node in FIG.~\ref{fig:InteractingMoose}.  
The factor of the perturbed theory including the weakly gauged group
is generically of the form given in the following table:

\begin{figure}
\includegraphics[scale=1]{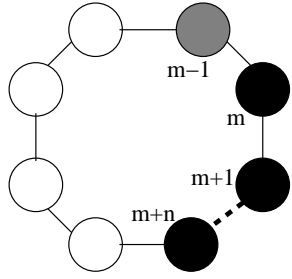}
\caption{\label{fig:InteractingMoose} Type 1 fixed point with chain of
$n+1$ interacting gauge groups.  The labels $(m-1),\dots,(m+n)$ index
the SU$(N)$ gauge group factors.  The gauge coupling for SU$(N)_{m-1}$ is
turned off at the fixed point.  Its $\beta$-function near the fixed point 
determines stability of the fixed point to development of the SU$(N)_{m-1}$
gauge coupling.}
\end{figure}

The node in the moose diagram adjacent to the last interacting node in the
chain at either end 
has only a single neighboring interacting node.  
For large $N$, near the 
edge of the conformal window where $N_f=N(2-\epsilon)$ with $\epsilon\ll 1$,
the fixed points are perturbative, so that the anomalous dimensions
can be assumed to be small \cite{Banks-Zaks}.  
In that case, the beta function for the $m$th interacting 
SU$(N)$ gauge group in FIG.~\ref{fig:InteractingMoose} 
is proportional to (assuming there is no accidental symmetry), \begin{eqnarray}
\beta_m^{num}
&=&-3G+\sum_i \mu_i(1-\gamma_i) \label{eq:betanum} \\
&=&-3N+\frac{N+2N_f}{2}\left(1-\gamma_{F_m}\right)+\frac{N}{2}
\left(1-\gamma_{Q_{m}}\right),
\end{eqnarray}
which is required to vanish at the fixed point.
On the other hand, if we now weakly gauge the gauge group at the $(m-1)$th
node, then the beta function for that gauge coupling is proportional
to,
\begin{eqnarray}
\beta_{m-1}^{num}
&=&-\left(3N-\frac{\left(N+2N_f\right)}{2}-
\frac{N}{2}\left(1-\gamma_{F_m}\right)\right) \nonumber \\
&=&\beta_{m}^{num}+
N_f \gamma_{F_m}+\frac{N}{2}\gamma_{Q_{m}}.  \label{eq:betam}
\end{eqnarray}

The anomalous dimensions can be calculated perturbatively, and at one loop are
given by, \begin{eqnarray}
\gamma_{F_i}&=&-\frac{(g_i^*)^2}{8\pi^2}\frac{(N^2-1)}{N}
\nonumber \\
\gamma_{Q_i}&=&-\frac{\left((g_i^*)^2+(g_{i+1}^*)^2\right)}{8\pi^2}\frac{(N^2-1)}{N},
\label{eq:pertanomdim}
\end{eqnarray}
where $g_i^*$ is the value of the SU$(N)_i$ gauge coupling at the fixed point.
The one-loop anomalous dimensions are negative, so that 
from (\ref{eq:betam}) at the fixed point
where $\beta_m=0$, we also have $\beta_{m-1}<0$.  This implies that in 
the perturbative
regime the Type 1 fixed point is unstable to turning on additional gauge 
couplings.  

Unfortunately, a simple argument similar to the above fails for the generic  
situation analogous to the B2 type fixed point, in which there is a gap
by one node between two chains of interacting gauge groups.  
However, it is natural to 
conjecture that such
fixed points, and indeed all fixed points in which not all couplings are turned
on, are always unstable.  We have not found an example to
the contrary in our limited survey to follow, with or without a brane.

\subsection{Will the Braneless Moose Hop After it Runs?}
We have argued that generically the braneless moose prefers to flow to IR
fixed points in which all of the gauge couplings are turned on.  This is
good from the perspective of deconstructed extra dimensions, because it
leads us to suspect that the hopping terms for the gauge fields are 
relevant operators and survive RG running down to 
scales for which the theory is supposed to
appear extra dimensional.  

However, we have not yet studied the hopping potential for the $N_f$
bulk flavors.  The hopping superpotential contains two types of operators:
$\Fbar_i F_i$, which is relevant at lowest order in the couplings; and $\Fbar_i
Q_i F_{i+1}$, which is marginal at lowest order.  The latter operators are 
dangerous,as quantum
corrections may easily 
make these operators irrelevant.  If that were the case, then
the hopping superpotential for the flavors would not exist at energies much 
below $\Lambda_{QCD}$, the strong coupling 
scale for each SU$(N)$ gauge group factor,
which is assumed to be larger than the scale of the highest KK mass so that
our fixed point analysis is valid (and because
the latticized theory requires strong coupling at that scale).  We study the
relevance of the hopping superpotential in this section.

The first comment to be made is that the superpotential is not made irrelevant
by wavefunction renormalization at one-loop.  This is because the anomalous
dimensions of the flavors and link fields are all negative at one loop.
At this order, the anomalous dimensions are given by 
(\ref{eq:pertanomdim}), where all of the gauge couplings are equal by
virtue of the $\mathbb{Z}_k$ symmetry of the fixed point.
From these anomalous dimensions we determine the R-charges of the fields
and the operator $\Fbar_i Q_i F_{i+1}$
using (\ref{eq:R-anomdim}):
 \begin{eqnarray}
r_F&=&\frac{2}{3}+\frac{\gamma_F}{3}\simeq\frac{2}{3}-\frac{(g^*)^2}{8\pi^2}
\frac{(N^2-1)}{3N} \nonumber \\
r_Q&=&\frac{2}{3}+\frac{\gamma_Q}{3}\simeq\frac{2}{3}-
\frac{(g^*)^2}{4\pi^2}\frac{(N^2-1)}{3N} \nonumber \\
r_{\Fbar_i Q_i F_{i+1}}&\simeq&2-
\frac{(g^*)^2}{2\pi^2}\frac{(N^2-1)}{3N} \ <2. \end{eqnarray}
Hence, we see explicitly that the R-charge of the superpotential is less than
2 perturbatively, 
and hence the hopping terms are relevant deformations near the
perturbative fixed point with vanishing superpotential.

To address the issue nonperturbatively we use the result of
a-maximization (\ref{eq:Zk-R}).  
The question is whether
or not the R-charges at the stable fixed point satisfy $r_{\Fbar_i}+r_{Q_i}+r_{
F_{i+1}}\leq 2$.  We find,
\begin{eqnarray}
r_{\Fbar_i Q_i F_{i+1}}&=& 
2-\frac{3 N_f^2-6 N N_f+(2 N-N_f) \sqrt{20 N^2-N_f^2}}{6 N^2-3 N_f^2}.
\label{eq:Zk-rW}
\end{eqnarray}
One can check from this expression
that in the naive conformal window $N/2<N_f<2N$,
$r_{\Fbar_i Q_i F_{i+1}}<2$.  
At the upper edge of the conformal window, when $N_f=2N$, 
$r_{\Fbar_i Q_i F_{i+1}}=2$ as expected because that is where the IR
theory becomes free.  Hence, the hopping superpotential is 
a relevant deformation of the theory in the conformal window, 
and we conclude that the hopping 
superpotential will indeed survive the RG flow down to the scale of the highest
KK mass $\Lambda_{KK}$.

\subsection{$\mathbb{Z}_k$ Symmetric Fixed Point with Hopping Superpotential}
If we assume that the hopping superpotential does not eliminate the
interacting IR fixed point, 
then we can calculate the fixed point anomalous dimensions
of the fields in the presence of the superpotential.
We are still considering the origin of
moduli space, {\em i.e.} vanishing link field vevs, so strictly speaking
this theory describes an approximate fixed point at energies
much higher than $\Lambda_{KK}$.  Assuming we can neglect the mass terms in
the superpotential at this scale, the anomaly
freedom condition (\ref{eq:anom-nobrane}) is supplemented with the
condition that the superpotential terms $\Fbar_i Q_i F_{i+1}$
have vanishing $\beta$-function at the
fixed point, namely $2 r_F+r_Q = 2$ (assuming $r_F=r_{\overline{F}}$, which 
will not be true in the more general models which follow). 
Together with (\ref{eq:anom-nobrane}), this
determines
$r_F=1$ and $r_Q=0$, which implies that the mass term is in fact
marginal at the
fixed point, and for consistency with unitarity
we would also conclude that the operators
$(\prod_{i=1}^k Q_i)^n$ are free and have dimension 1. These operators are
related to Wilson loops around the extra dimension, 
and the interpretation of the fact that those 
operators are free above the scale of the highest KK mass
deserves exploration.

\section{The Moose With a Brane}
\label{sec:brane}
In this section we add additional flavors localized to one lattice site, which
we label the first lattice site and refer to as the brane.  
We still consider the circular moose model 
given by the table (\ref{moosetable}).  The
result of a-maximization in this case is sensitive to the number of lattice
sites, and becomes numerically unwieldy with more than a few lattice sites.
All analytic expressions in this section are valid when there is no
accidental symmetry.
We numerically study some examples with two, 
three and four lattice sites, and give more general results where possible.  
In the numerical examples here and in the following section we account for
all accidental symmetries that are required by unitarity.

As before we consider the circular moose model in the naive conformal
window $N/2 < N_f < 2N-N_b$.  To analyze the stability of the fixed
points we will use the same techniques as with the braneless moose. We only
analyze fixed points whose structure and stability analysis differs
from the braneless case. Type
A fixed points for the moose with a brane, {\em i.e.} fixed points with a 
single nonvanishing SU$(N)_i$ gauge coupling,  can be divided into two types
which we call
type A1 and type A2. A type A1 fixed point is a fixed point with the
brane gauge coupling nonvanishing. A type A2 fixed point
is a fixed point with the
gauge coupling next to the brane nonvanishing ($g_2$ or $g_k$). At an
A1 type fixed point, the effective theory is like SQCD with
$(N+N_f+N_b)$ flavors. The R-charges are given by

\begin{equation}
r= \frac{N_f+N_b}{N_f+N_b+N}
\label{eq:rb}\end{equation}
Perturbing the fixed point by weakly gauging $SU(N)_2$ we get the beta function, for two sites
\begin {eqnarray} \label{eq:beta2I}
\beta_2^{I}&=&-\frac{g_2^3}{16\pi^2}\frac{2N_fN+3N_bN-N_f^2-N_fN_b}{\left(N+N_f+N_b\right)\left(1-\frac{g_2^2N}{8\pi^2}\right)}
\end {eqnarray}
and for more than two sites
\begin{eqnarray}\label{eq:beta2II}
\beta_2^{II}&=& -\frac{g_2^3}{32\pi^2}\frac{2N^2+3NN_f-2N_f^2+5N_bN-2N_bN_f}{\left(N+N_f+N_b\right)\left(1-\frac{g_2^2N}{8\pi^2}\right)}
\end{eqnarray}
The beta functions in (\ref{eq:beta2I}) and (\ref{eq:beta2II}) are
positive in the
whole  naive conformal window.  Hence, Type A1 fixed points are
always unstable to development of neighboring gauge couplings. 
The Euler anomaly at the type A1 fixed point is given by:\begin{eqnarray}
a_1^*&=&\frac{1}{48}\left[k(10N^2+2NN_f-9)-N/(\left(N+N_f+N_b\right)^2)\left(N\left(20N^2-14NN_b-12NN_f-7N_b^2\right.\right.\right.\nonumber\\
&&\left.\left.\left.-10N_fN_b-3N_f^2\right)+2N_f^3+4N_f^2N_b+2N_fN_b^2\right)\right]
\end{eqnarray}

\begin{figure}

\includegraphics[scale=0.5]{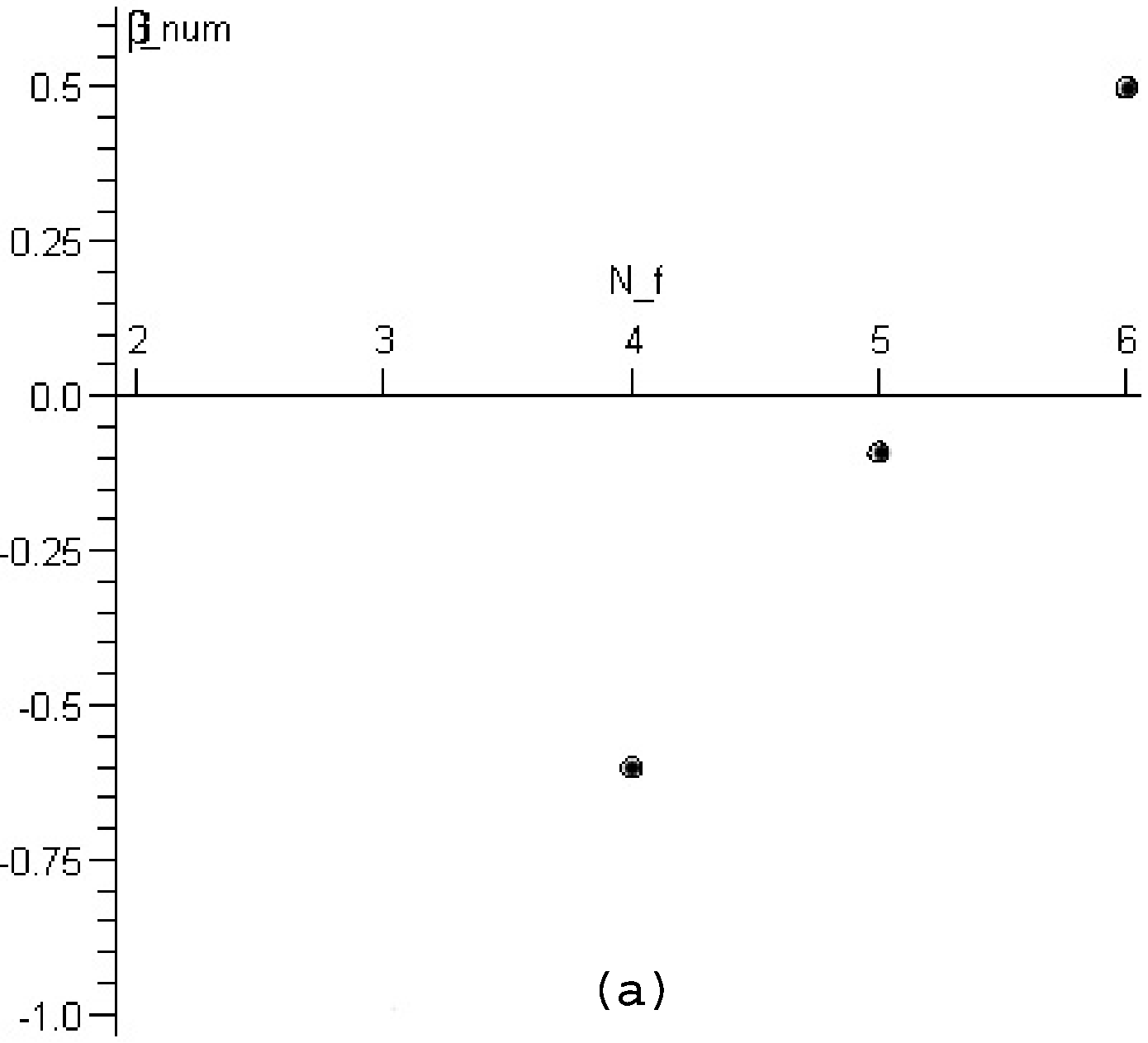}
\hfill
\includegraphics[scale=0.5]{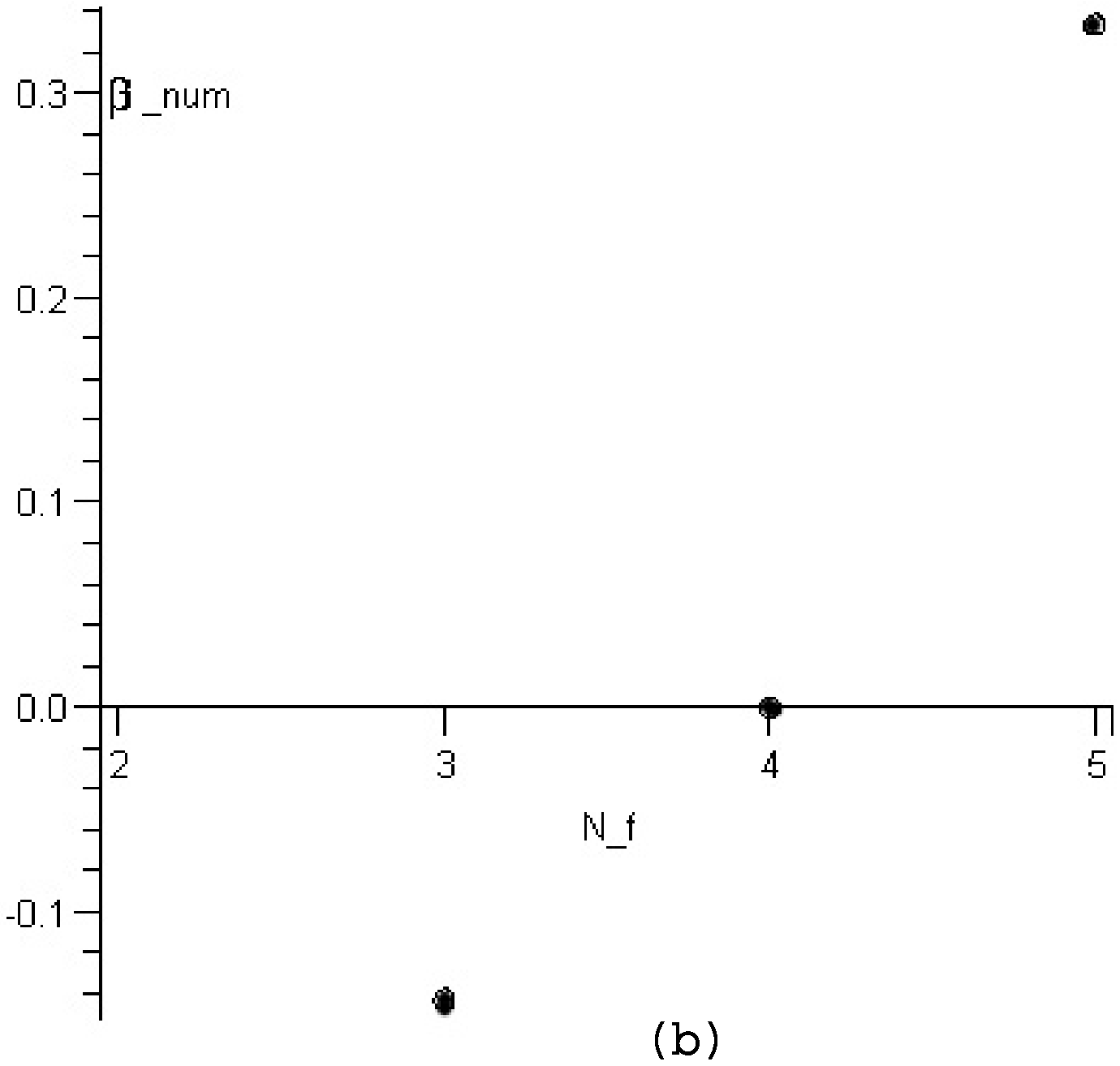}
\caption{\label{betaN4Nb3}Numerator of the beta function (\ref{eq:betanum})
for (a) $k>2$, $N=6$, $N_b=5$; and (b) $k=2$, $N=4$, $N_b=2$.}
\end{figure}

Type A2 fixed point the R-charges is the same as equation (\ref{eq:r}), the NSVZ beta function for two sites with turning on weakly the $g_1$  is given by
\begin{equation}
\beta_1^{I} = -\frac{g_1^3}{16\pi^2}\frac{2NN_f-N_f^2-N_bN-N_bN_f}{\left(N+N_f\right)\left(1-\frac{g_1^2N}{8\pi^2}\right)}
\label{braneA2I}
\end{equation}
for more than two sites
\begin{equation}
\beta_1^{II} = -\frac{g_1^3}{32\pi^2}\frac{2N^2+3NN_f-2N_f^2-2N_bN-2N_bN_f}{\left(N+N_f\right)\left(1-\frac{g_1^2N}{8\pi^2}\right)}
\label{braneA2II}
\end{equation}
From FIG.~\ref{betaN4Nb3}b in the two-site model we see that
this type of fixed point
could be stable or unstable depending on the number of flavors and
colors, as was observed in \cite{Barnes}; 
this behaviour is also true for more than two sites, as can
be see from FIG.~\ref{betaN4Nb3}a. In the two site case, if this type
fixed point is stable, there is a flow from the A1 type fixed point to the A2
type fixed point. When this type fixed point is unstable the flow will
be from A1 and A2 to a fixed point where both gauge couplings are turned
on.  We find that for $N_b < 1/2(7-2\sqrt{6})N$, with more than two lattice
sites the A2 type fixed point will be stable for
\begin{equation}
N_f > \frac{3}{4}N-\frac{1}{2}N_b+\frac{1}{4}\sqrt{25N^2-28N_bN+4N_b^2}
\end{equation}
when $N_b > 1/2(7-2\sqrt{6})N$ this kind of fixed point will be always
stable from $g_1$ pertubation, because the beta function will be
always positive in the whole conformal window. The $a$ for this kind of fixed points is
\begin{eqnarray}
a_2^* &=&\frac{1}{48\left(N+N_f\right)^2} \left[k\left(10N^4+22N^3N_f+14N^2N_f^2-9N^2-18NN_f-9N_f^2.+2NN_f^3\right)\right.\nonumber \\
  && \left. -20N^4+12N^3N_f+3N^2N_f^2+2N_bN^3+4N_bN^2N_f+2NN_bN_f^2-2NN_f^3\right]
\end{eqnarray}

For fixed points with two nonvanishing gauge couplings, we will consider four
different cases.  ({\bf a}) First we consider a case when sites 2 and 3 
(or $k-1$ and $k$) turn on. Perturbing the fixed point by the SU$(N)_1$ 
coupling, the effective theory for this type of fixed
point is as follows:
\begin{eqnarray}
&&\begin{array}{c|ccc|c}
& SU(N)_1 & SU(N)_2 & SU(N)_3 & U(1)_R\\
\hline
Q_1 &\Yfund & \overline{\Yfund} & 1 & r_F\\
Q_2 & 1 & \Yfund &\overline{\Yfund} & r_2\\
(N+N_f)F_3 & 1 & 1 & \Yfund & r_F \\
N_f \overline{F}_3 & 1 &1 & \overline{\Yfund} & r_F \\
N_f F_2 &1 & \Yfund &1 & r_F\\
N_f \overline{F}_2 & 1 & \overline{\Yfund} & 1 & r_F \\
(N_f+N_b+N)\overline{F}_1 & \overline{\Yfund} & 1 & 1 & 2/3 \\
(N_f+N_b)F_1 &\Yfund & 1 &1 & 2/3\\
\hline
\end{array}
\end{eqnarray}
The $R$ charges for this kind of fixed point is the same as in equation 
(\ref{eq:rF1}) for $r_F$ and equation (\ref{eq:rQ}) 
for $r_2$. The beta function by perturbing the brane 
gauge coupling for $k=3$ is,
\begin{eqnarray}
\beta_1^{BI}
&=&-\frac{g_1^3}{16\pi^2}\frac{1}{(N^2-4NN_f-4N_f^2)\left(1-\frac{g_1^2N}{8\pi^2}\right)}\left(9N^3-N^2N_f-8NN_f^2+4N_f^3-N_bN^2
\right.\nonumber\\ &&\left.+4N_bNN_f -N^{2}\sqrt{73N^2-4NN_f-4N_f^2}
+4N_bN_f^2\right)
\end{eqnarray}
For $k>3$ the beta function is equal to
\begin{eqnarray}
\beta_1^{BI'}&=&-\frac{g_1^3}{32\pi^2}\frac{1}{(N^2-4NN_f-4N_f^2)\left(1-\frac{g_1^2N}{8\pi^2}\right)}\left(11N^3-10N^2N_f-12NN_f^2+8N_f^3
\right.\nonumber\\ &&\left.-2N_bN^2 +8N_bNN_f-N^{2}\sqrt{73N^2-4NN_f-4N_f^2}
+8N_bN_f^2\right)
\end{eqnarray}
The beta function for this kind of fixed point could be negative or
positive depending on the number of flavors.
We also could perturb the fixed point with nonvanishing coupling $g_4$ for this
case and the beta function is the same as equation (\ref{eq:beta3B1}). From
the braneless case we already know that this kind fixed point is unstable 
to such a perturbation. The expression for the value of $a$ at
this type of fixed point is,
\begin{eqnarray}
a_a^* &=&\frac{1}{48(N^2-4NN_f-4N_f^2)^2}\left(N^6\left(-651+10k\right)+N
\left(32N_bN_f^4+32kN_f^5-64N_f^5-288kN_f^3\right)\right.\nonumber\\
  &&+N^5(-78N_fk+2N_b+73\sqrt{73N^2-4NN_f-4N_f^2}-916N_f)-144kN_f^4\nonumber\\
&&+N^4(-9k+64kN_f^2+142N_f\sqrt{73N^2-4NN_f-4N_f^2}-16N_bN_f-640N_f^2)\nonumber\\
&&+N^3(72N_fk-12N_f^2\sqrt{73N^2-4NN_f-4N_f^2}+16N_bN_f^2+336kN_f^3+448N_f^3)\nonumber\\
&&\left.+N^2(-8N_f^3\sqrt{73N^2-4NN_f-4N_f^2}+224kN_f^4+112N_f^4+64N_bN_f^3-72kN_f^2)\right)
\label{eq:aa}\end{eqnarray}

\begin{figure}
\includegraphics[scale=0.5]{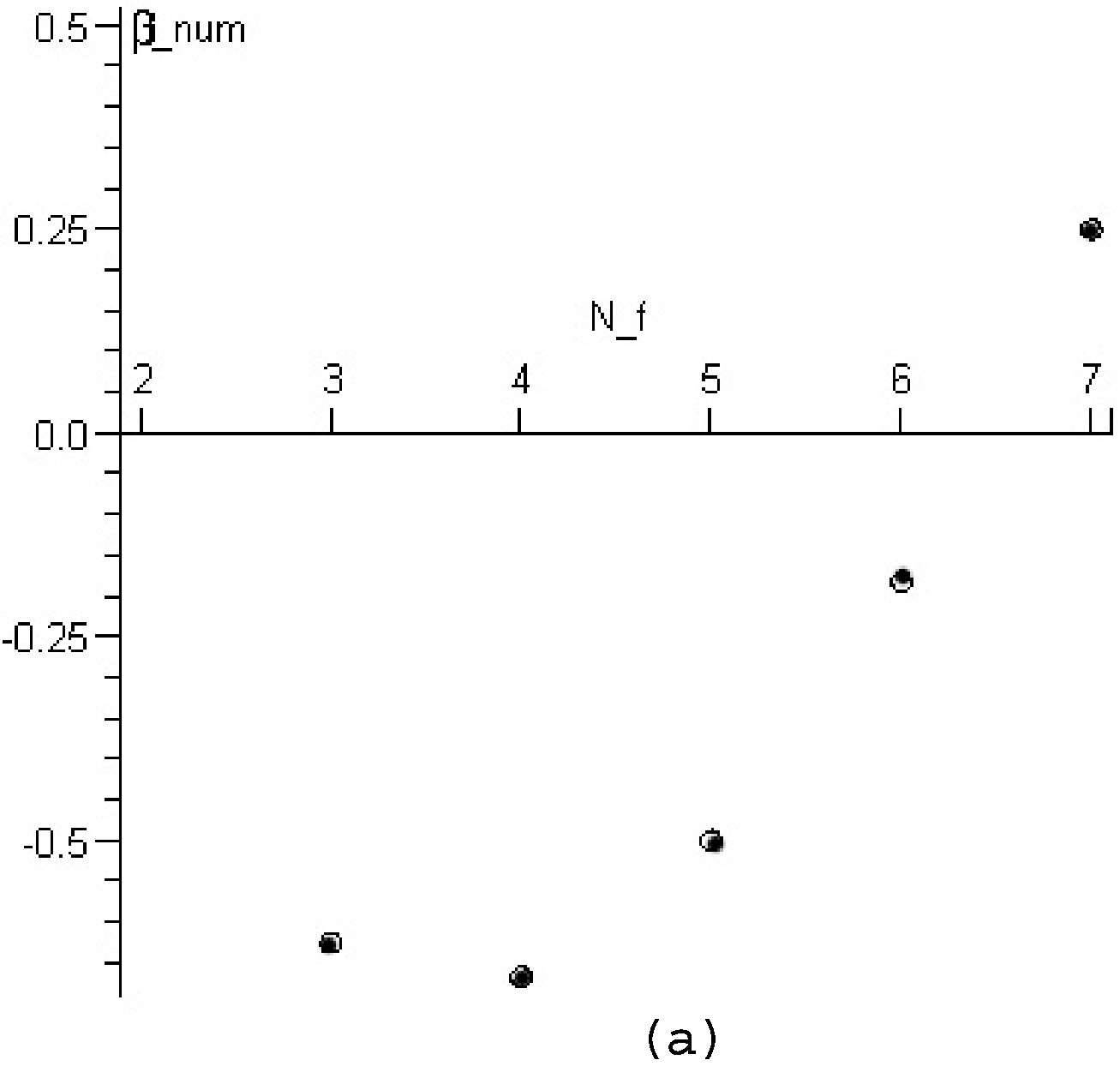}
\hfill
\includegraphics[scale=0.5]{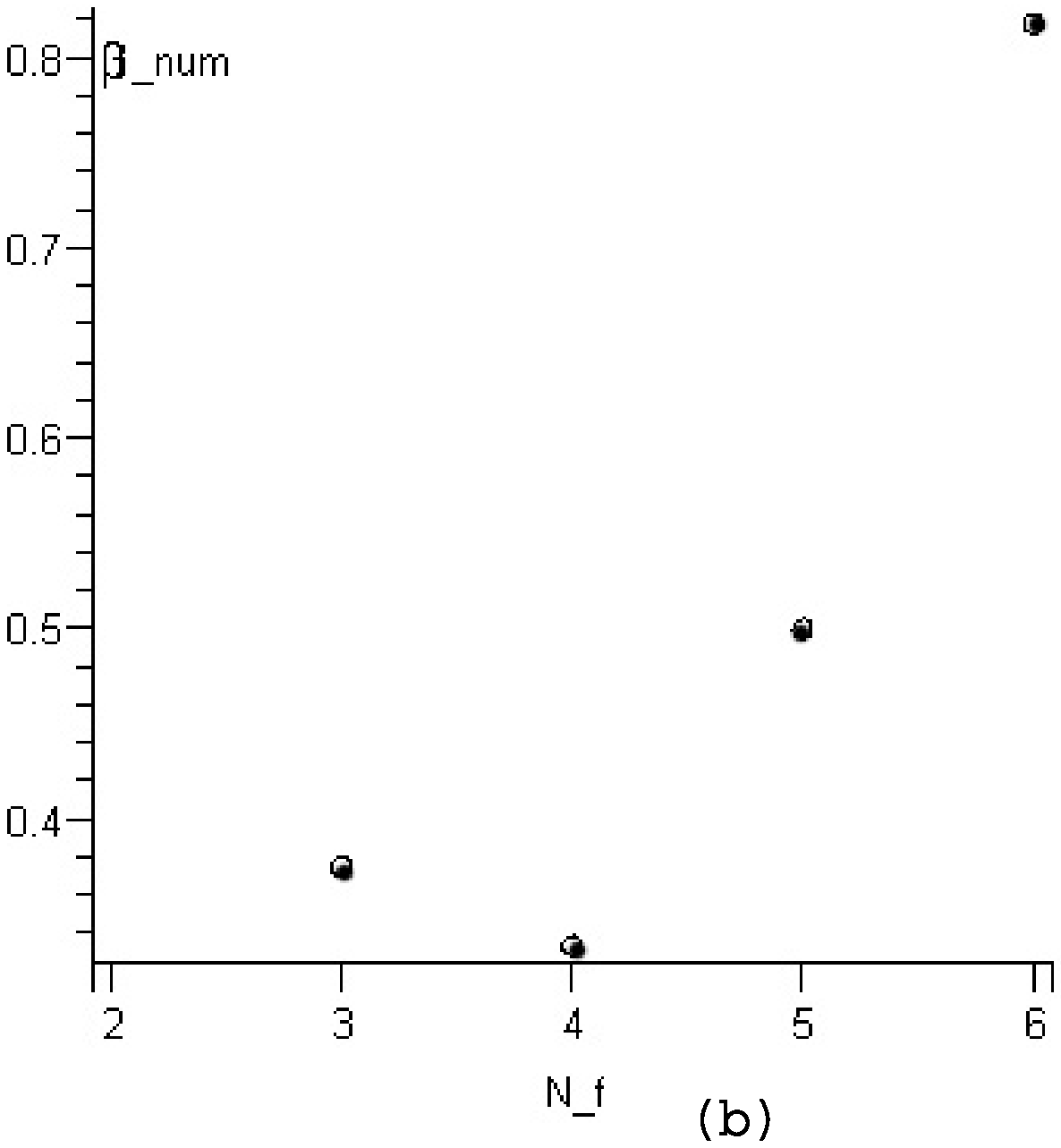}
\caption{\label{typeA2} Numerator of the beta function (\ref{eq:betanum}). 
(a) for $N=5$, $N_b=2$; and (b) for $N=5$, $N_b=3$.}
\end{figure}

({\bf b}) The fixed point with $g_k$ and $g_2$ turned on
is like the type B2 fixed point for braneless moose. 
The effective theory for this type of fixed point, perturbed by SU$(N)_1$, is:
\begin{eqnarray}
&&\begin{array}{c|ccc|c}
  & SU(N)_k & SU(N)_1 & SU(N)_2 & U(1)_R \\
\hline
  Q_{k} & \Yfund  & \overline{\Yfund}  & 1 & r \\
  Q_1 & 1 & \Yfund  & \overline{\Yfund} & r \\
N_f\ F_k & \Yfund & 1  &1 & r \\
(N_f+N)\  \overline{F}_k & \overline{\Yfund} & 1 &1 & r \\
(N_f+N_b)\  F_1 & 1 & \Yfund  &1 & 2/3 \\
(N_f+N_b)\  \overline{F}_1 & 1 & \overline{\Yfund} &1 & 2/3 \\
(N_f+N)\  F_2 & 1 & 1& \Yfund  & r\\
N_f\  \overline{F}_2 & 1 & 1  &\overline{\Yfund} & r \\

\hline
\end{array}
\end{eqnarray}
The $r$ is the same as equation (\ref{eq:r}), this type of fixed points only occur for number of sites $k>3$, because for $k=3$ this fixed point will be
equivalent to case ({\bf a}). The beta function of the weakly gauged SU$(N)_1$ is,
\begin {equation}
\beta_1^{BII} = -\frac{g_1^3}{16\pi^2}\frac{N_f\left(2N-N_b-N_f\right)-N_bN}{\left(N+N_f\right)\left(1-\frac{g_1^2N}{8\pi^2}\right)}
\end {equation}
The sign of the beta function depends on $N_b$ and $N_f$. We find numerically
that for $N_b > 0.584 N$ the beta function is positive in the whole 
naive conformal window. An example is shown in FIG.\ref{typeA2}. 
We also could give small pertubation to $g_3$ or $g_{k-1}$. The beta function for $k=4$ is the same as equation (\ref{eq:beta2B2}), so for $k>4$ the beta function is
\begin{equation}
\beta_{k-1}^{BII}=-\frac{g_{k-1}^3}{32\pi^2}\frac{2N^2+3NN_f-2N_f^2}{(N+N_f)\left(1-\frac{g_{k-1}^2N}{8\pi^2}\right)}
\end{equation}
We find that for $N_f < 2 N$ the fixed points is always unstable. The value of $a$ for this fixed points can be written as follows:
\begin{eqnarray}
a_{BII}^*&=&\frac{1}{48(N+N_f)^2}\left(-18kNN_f+2NN_f^3k+13N_bN^2N_f+2NN_bN_f^2+14N_f^2kN^2\right.\nonumber\\
&&\left. -40N^4+24N_fN^3+6N^2N_f^2-4NN_f^3+10N^4k+11N_bN^3-9kN^2
\right. \nonumber \\
&&\left.-9kN_f^2+22N_fkN^3\right)
\end{eqnarray}

({\bf c}) In the third case the couplings $g_1$ and $g_3$ are non vanishing. 
We will study this case for $k=3$ in the next example, so 
we now consider $k>3$.
The effective theory for this type fixed point perturbed by SU$(N)_2$ is
given by the following table: 
\begin{eqnarray}
&&\begin{array}{c|ccc|c}
  & SU(N)_1 & SU(N)_2 & SU(N)_3 & U(1)_R \\
\hline
  Q_{1} & \Yfund  & \overline{\Yfund}  & 1 & r_1 \\
  Q_2 & 1 & \Yfund  & \overline{\Yfund} & r_2 \\
(N_f+N_b)\ F_1 & \Yfund & 1  &1 & r_1 \\
(N_f+N+N_b) \overline{F}_1 & \overline{\Yfund} & 1 &1 & r_1 \\
(N_f)\  F_2 & 1 & \Yfund  &1 & 2/3 \\
(N_f)\  \overline{F}_2 & 1 & \overline{\Yfund} &1 & 2/3 \\
(N_f+N)\  F_3 & 1 & 1& \Yfund  & r_2\\
N_f\  \overline{F}_3 & 1 & 1  &\overline{\Yfund} & r_2 \\

\hline
\end{array}
\end{eqnarray}
the $R$ charges is the same as equation (\ref{eq:r}) for $r_2$ and equation 
(\ref{eq:rb}) for $r_1$. The beta function by small pertubation of $g_2$ is
\begin{eqnarray}
\beta_2^{BIII}=-\frac{g_2^3}{32\pi^2}\frac{4N^2N_f+3N^2N_b+2NN_f^2+4NN_fN_b-2N_f^3-2N_f^2N_b}{(N+N_f)(N+N_f+N_B)\left(1-\frac{g_2^2N}{8\pi^2}\right)}
\end{eqnarray}
\begin{figure}
\includegraphics[scale=0.55]{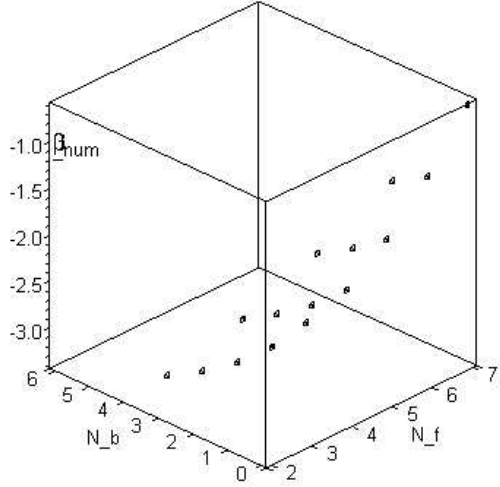}
\caption{\label{3d13} Numerator of the beta function (\ref{eq:betanum}) 
for type BIII fixed points with $N=4$.}
\end{figure}

We could see the behavior of the beta function from
FIG.~\ref{3d13}. We see that this type
of fixed point is always unstable under $g_2$ pertubation. For $k>4$ if we
make a small pertubation on $g_k$ or $g_4$, the stability analysis is 
the same as the case of one gauge coupling turned on.

({\bf d}) The last type of fixed point with two gauge couplings 
turned on has $g_1$ and either $g_2$ or $g_k$ nonvanishing. 
We studied this case numerically, and we give some examples for
$N = 4$ and $k=3$ with $g_1$ and $g_2$ turned on, perturbed by $g_3$.  
We study the three-site model in full detail in the next section.
\begin{table}[htbp]
\caption{Type four 2 gauge coupling turn on for $k=3$, $N=3$. An asterisk 
indicates that the fixed point is absent in that theory.}
\begin{tabular}{c|c|c|c|}
$N_b$ & $N_f$ & $a$ & $\beta_3^{num}$\\
\hline
1 & 2 & 5.0456 & -2.3774\\
1 & 3 & 5.9347 & -2.0058\\
1 & 4 & 6.5617 & -1.4758\\
2 & 2 & 5.2819 & -2.6383\\
2 & 3 & 6.1077 & -2.2071\\
3 & 2 &* &*\\
\hline 
\end{tabular}
\label{B13}
\end{table}

In the examples in table (\ref{B13}) we see that this type of fixed
points is unstable in the whole naive conformal window.  In some examples
this type of fixed point doesn't
exist. This is because from the beta function analysis those points would
be saddle points in the RG flow, and both a perturbative analysis and
consistency with the a-theorem suggest that those fixed points are absent.
We will discuss the absence of these fixed points in more detail in the
next section.  From
our discussion we could now do some example cases for $k=3$,
$N=3$. Numerically we see in these examples that the 
stable fixed points always have the
smallest value of $a$. In general we find two types of stable fixed
points, the first one having all the gauge couplings turned on and 
the other one
having only the brane gauge coupling turned off. 
In table (\ref{examp}) A, B and C
tells how many couplings are turned on (1, 2 and 3, respectively), 
and the subcript tells at which
sites the couplings are turned on. $\beta_i^{num}$ is given by
(\ref{eq:betanum}).  The three examples in the table demonstrate
the three different types of RG flow patterns that we find.  We will 
describe the generic flows in more detail in the next section.
\begin{table}[htbp]
\caption{Fixed points and stability For $N=3$}
\begin{tabular}{cc|c|c|c|c|c}
&$Type$ & $a$ & $\beta_1^{num}$ & $\beta_2^{num}$ & $\beta_3^{num}$ &remark\\
\hline
$N_b=3$ $N_f=2$ & $A_1$ & 6.163 &  & -3.812&-3.812&\\
&$A_2$ & 5.427 & 2.00 & & -2.80 & same as $A_3$\\
& $B_{12}$ & 5.426  &  & &-2.835 & doesn't exist (same as $B_{13}$)\\
& $B_{23}$ & 4.999 & 0.982 & & & stable\\
& $C_{123}$ &4.971  &  & & & doesn't exist\\
\hline
$N_b=2$ $N_f=2$ & $A_1$ & 5.945 &  & -3.571&-3.571&\\
&$A_2$ & 5.3025 & -0.800 & & -2.8 & same as $A_3$\\
& $B_{12}$ & 5.28195  &  & &-2.64 & same as $B_{13}$ \\
& $B_{23}$ & 4.8739 & -0.0179 & & & \\
& $C_{123}$ &4.8738  &  & & & stable\\
\hline
$N_b=2$ $N_f=3$ & $A_1$ & 6.4130 &  & -2.81&-2.81&\\
&$A_2$ & 6.1094 & -0.25 & & -2.25 & same as $A_3$\\
& $B_{12}$ & 6.1077  &  & &-2.207 & same as $B_{13}$\\
& $B_{23}$ & 5.9061 & 0.259 & & & stable \\
& $C_{123}$ &5.9042  &  & & & doesn't exist \\
\hline
\end{tabular}
\label{examp}
\end{table}

\begin{figure}[htbp]
\includegraphics[scale=0.5]{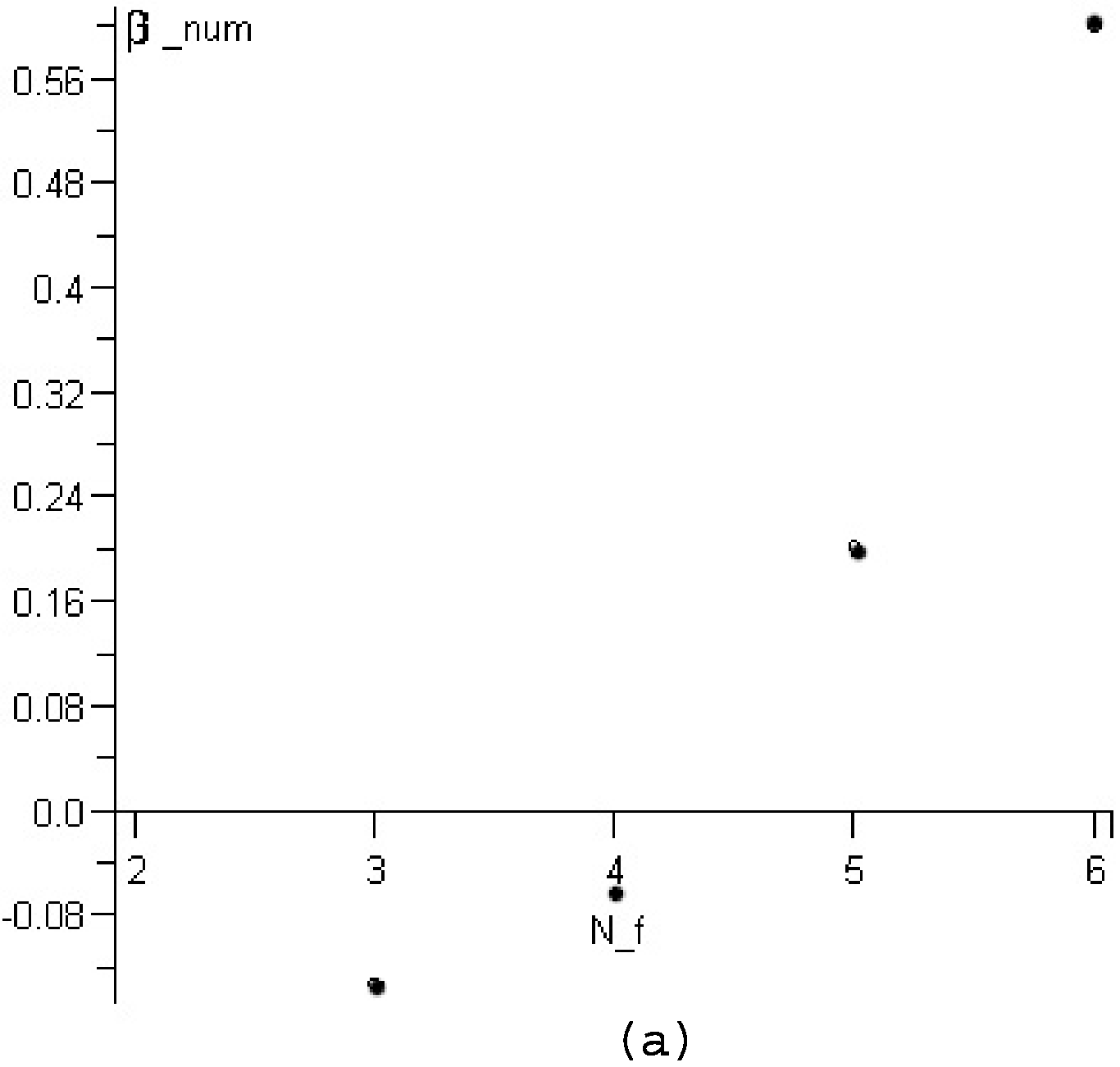} \hfill
\includegraphics[scale=0.5]{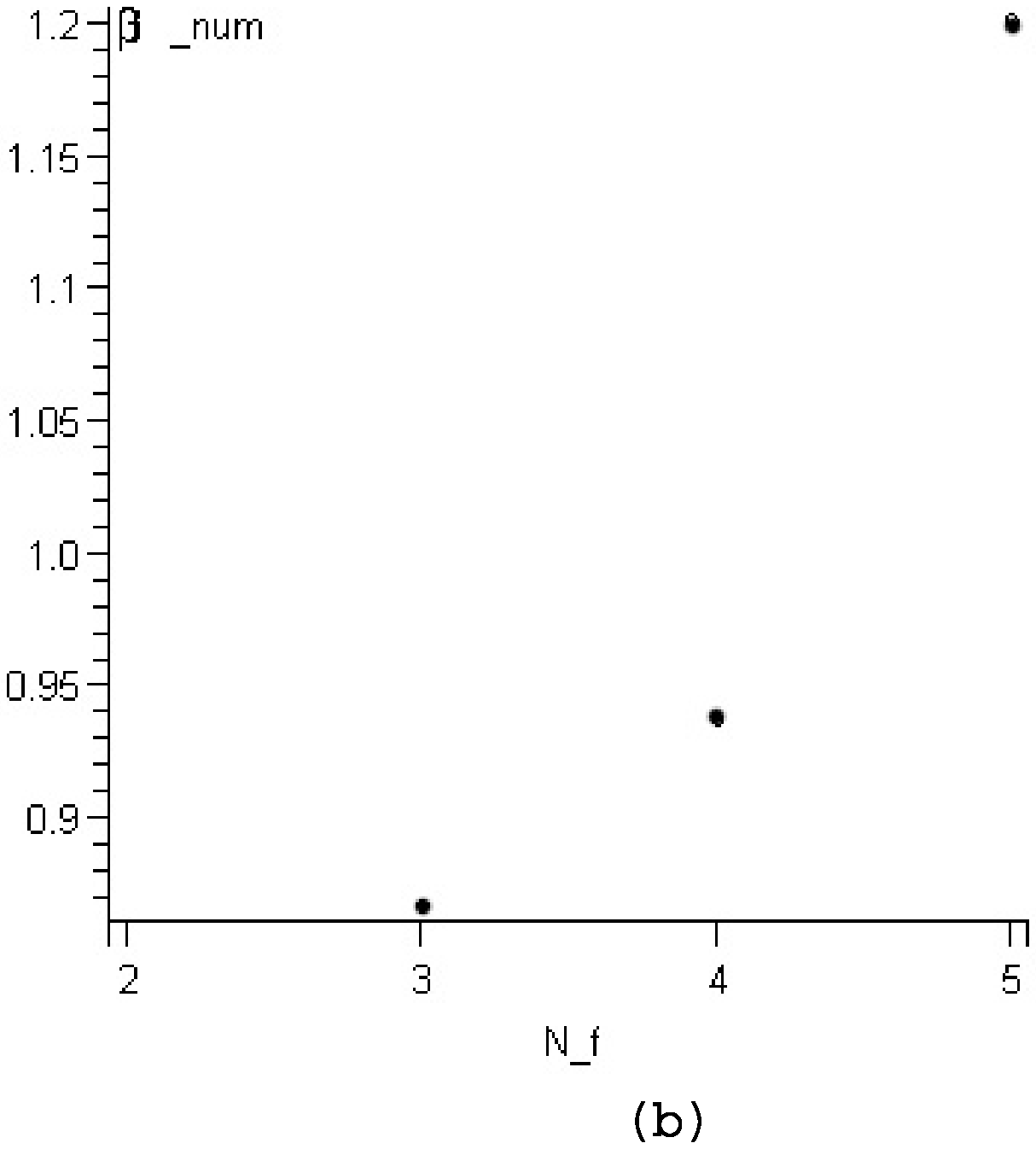}
\caption{\label{3coupling}Numerator of the beta function (\ref{eq:betanum}) 
for $k=4$, $N=5$,(a) $N_b=3$ and (b) $N_b=4$.}
\end{figure}

The fixed point stability analysis in theories with more than three gauge
group factors is similar.  In the theory with four lattice sites we plot the 
numerator of the SU$(N)_1$ beta function near the fixed point with the 
remaining gauge couplings turned on in FIG.~\ref{3coupling}.
Once again, we find that whether or not the brane gauge coupling is 
nonvanishing at the stable fixed point depends on $N_b$ and $N_f$.

\begin{figure}[htbp]
\includegraphics[scale=0.5]{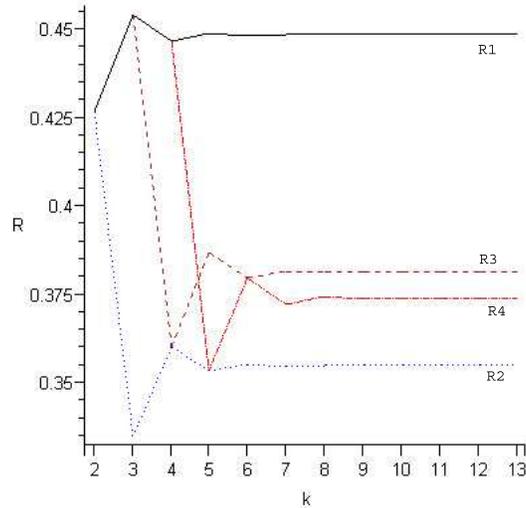}
\caption{\label{rQgraph}Values of $r_{Q_i}$ for $N =4$, $N_f=3$, $N_b=2$, 
at the stable fixed point as a function of the number of lattice sites $k$.}
\end{figure}

While we do not have analytic solutions for the dependence of the R-charges
on the number of lattice sites, it is interesting
to note that they do have smooth asymptotic behavior as $k\rightarrow\infty$.
FIG.~\ref{rQgraph} demonstrates the variation of the R-charges for the first
few link fields near the brane as a function of $k$ for the moose models with 
$N=4$, $N_f=3$ and $N_b=2$ at the fixed point with all gauge couplings
turned on.  

\begin{figure}[htbp]
\includegraphics[scale=0.5]{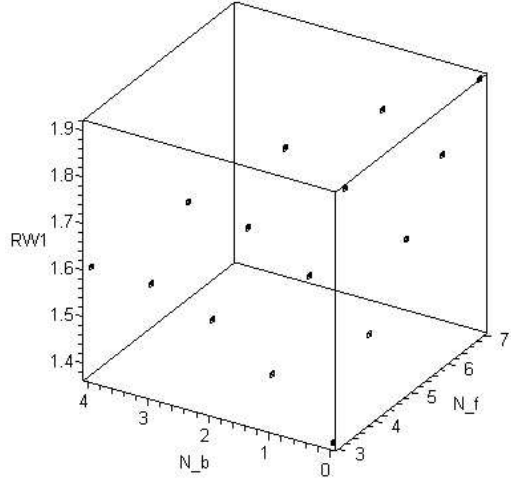}
\caption{\label{r_W}R-charge of the operator $\overline{F}_1Q_1F_2$
at the stable fixed points in the three-site model, $N=4$.}
\end{figure}

We would still like to see that the hopping superpotential is a relevant 
deformation of the fixed point theory.
FIG.~\ref{r_W}
demonstrates the variation of the dimension of the operator $\overline{F}_1
Q_1 F_2$ on $N_f$ and $N_b$ for fixed $k=3$, $N=4$.   Note
that the operator remains relevant for the whole naive coformal window.  
We have checked numerically that in the three-site models in their
conformal window with $N=2,3,4,5$,  each operator in the 
hopping superpotential is relevant at the stable fixed point.  This result
is heartening from the perspective of the extra dimensional interpretation,
although we do not have any general conclusions of this sort in the 
continuum limit.

\section{The A-theorem}
Cardy has suggested that the Euler anomaly $a$ decreases monotonically
from the UV to the  IR, and is everywhere positive.  The conjecture that
$a$ plays the role of a Zamolodchikov $C$-function in four dimensions
is known as the a-theorem.  At conformal fixed
points the value of $a$ is given by the a-maximization procedure of
\cite{a-max} used above.  One can consider flows from unstable to
stable fixed points of the type studied above to test Cardy's conjecture.  

The simplest example of such an RG flow is from the free theory 
to a fixed point in which a single gauge coupling is
turned on.  A large set of flows of this sort in theories with various
matter fields, gauge groups and superpotentials was studied in \cite{Freedman},
Focusing on the gauge group which is turned on, the change in $a$ from
the UV to the IR is given by \begin{eqnarray}
\Delta a &=& \frac{3(N+N_f) N}{16} \left(\frac{2}{9} + 3\left(\frac{N}{N_f+N}\right)^3
-\frac{N}{N_f+N}\right), \end{eqnarray}
which one can check is positive in the conformal window $N/2<N_f<2N$.

An analytical
analysis of $\Delta a$ between fixed points with additional nonvanishing
couplings quickly becomes difficult.  Hence, we resort for now
to a numerical study for a range of $N,\ N_f$ and $N_b$.  
We have studied all of the fixed points for
the supersymmetric moose models with three gauge groups (the 
``three-site model'') in the naive conformal window 
$N/2<N_f<(2N-N_b)$ with $N = 2,3,4,5$.  
We used Mathematica to analytically 
solve for the R-charges which maximize $a$, and to account for accidental
symmetries where they were required for unitarity.
We numerically compared the
differences in the values of $a$ at fixed points with one, two or three
gauge couplings turned on.  

We label the fixed points as follows:
\begin{center}
\begin{tabular}{ccl}
A1 && Brane gauge coupling turned on \\
A2/A3 && Non-brane gauge coupling turned on \\
B12/B13 && Two gauge couplings turned on, including the brane \\
B23 && Two gauge couplings turned on, not including brane \\
C123 && All three gauge couplings turned on. \end{tabular}
\end{center}

The analytic solutions for the R-charges which maximize $a$ are complicated and
in themselves not enlightening.
Here we summarize
the conclusions of our analysis of the a-theorem.  We find that the
values of $a$ for each set of $N$, $N_f$ and $N_b$ follow the following
pattern: \begin{equation}
a_{\rm A1}\geq a_{\rm A2/A3}\geq a_{\rm B12/B13} \geq a_{\rm B23}\geq a_{\rm 
C123}, \label{eq:fixedpoints}\end{equation}
where fixed points which do not exist for a
given $(N,N_f,N_b)$ are to be removed from the above inequalities, which
are otherwise unchanged.  In each case we find a picture 
consistent with the a-theorem, 
and the flows are always in the direction, \begin{equation}
{\rm A1}\rightarrow{\rm A2/A3}\rightarrow{\rm B12/B13} \rightarrow
{\rm B23}\rightarrow{\rm C123}, \end{equation}
where again nonexistent fixed points are understood to be removed from 
the above pattern of RG flows.
The test for the direction of RG flow is from the $\beta$-functions of the
weakly coupled gauge groups in the neighborhood of the fixed points.

The $\beta$-functions for the bulk gauge couplings
at the A1 type fixed point are always negative, so that at least one of the
bulk gauge couplings will necessarily be turned on at the stable fixed point.
(In fact, both bulk gauge couplings will be turned on.)

The $\beta$-function for the additional bulk
gauge coupling at the A2/A3
type fixed point is always negative.  However, the $\beta$-function for
the brane gauge coupling near these fixed points 
is positive for \begin{eqnarray}
(N,N_f,N_b)&=&  
(3, 2, 3), \     
(4, 3, 4),\ 
(5, 3, 5),\  (5, 3, 6),\    
(5, 4, 5),\  
(5, 5, 4).\   
\nonumber \end{eqnarray}
Those fixed points are stable to perturbations by the brane gauge coupling,
and in those cases the brane gauge coupling is irrelevant in the IR.

The $\beta$-functions for the brane gauge coupling at the B23 type fixed 
points, {\em i.e.} with both bulk gauge couplings turned on, are positive
for all cases in which the brane gauge coupling is positive for the
A2/A3 type fixed point (enumerated above), in addition to the theories with,
\begin{eqnarray}
(N,N_f,N_b)&=&  
(3, 3, 2),\   
(4, 3, 3),
\  (4, 4, 3),\  (4, 5, 2),\  (5, 3, 4),\nonumber \\
&&  (5, 4, 4),\  (5, 5, 3),\  (5, 6, 3),
\  (5, 7, 2). \nonumber \end{eqnarray}

The $\beta$-functions for the bulk gauge couplings are all negative at the
B12/B13 type fixed points, so that {\em both of the bulk gauge couplings 
are turned on at each of the stable fixed points}, as expected. 

\begin{figure}
\includegraphics[scale=.7]{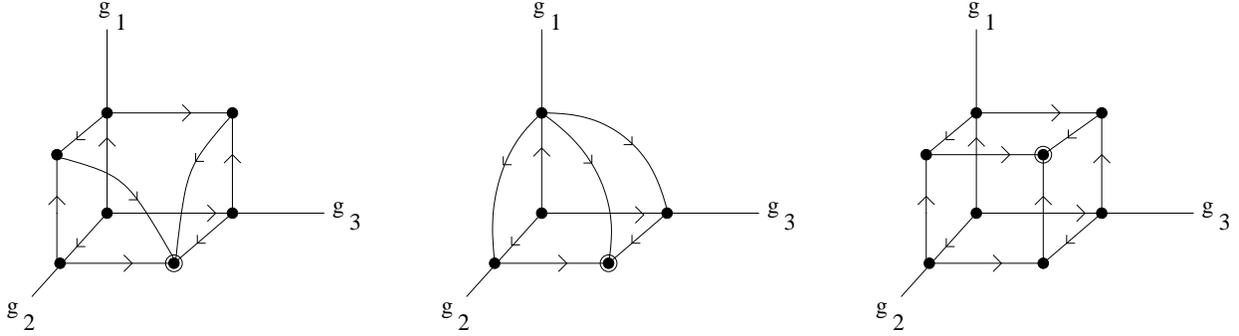}
\caption{\label{fig:RG}
The three classes of RG flows between fixed points in three-site models. The
gauge coupling at the ``brane'' is $g_1$.  The stable fixed point is circled
in each case.}
\end{figure}

To summarize, we have found 
that for each $(N,N_f,N_b)$ the RG flows between the fixed points
fall in one of three classes shown in Fig.~\ref{fig:RG}.
A few comments are in order here.  Note that there are examples of flows
for which the brane coupling is attractive if one of the bulk gauge couplings
is turned off, but becomes repulsive 
when the remaining bulk gauge coupling is turned on.  Furthermore, it is 
interesting to note that the a-theorem together with (\ref{eq:fixedpoints})
precludes the existence of certain saddle points 
in the RG flows which would allow flows like 
B12$\rightarrow$C123$\rightarrow$B23 or
A1$\rightarrow$B12$\rightarrow$A2.  Hence,
it follows that the a-theorem disallows otherwise
possible fixed points along the
trajectory from unstable to stable fixed points.  (At weakly coupled
fixed points an analysis such as in \cite{Barnes} should also demonstrate
that such saddle points are not allowed.)  As mentioned earlier,
we have also checked that all of the hopping terms in the superpotential
are relevant deformations of the stable fixed point theories in their absence
({\em i.e.} they have R-charge $<2$.)

There are many additional tests of the a-theorem which can
be done by studying flows between fixed points in other product group theories,
the simplest generalization of which is to include additional ``branes'' into
the analysis done here, {\em i.e.} varying the numbers of flavors at each
node of the moose.  We will not pursue a more general analysis here.

\section{Discussion}
We have shown that both the gauge couplings and the superpotential are 
generically relevant deformations of conformal fixed points
of moose models at which those couplings are turned off. The bulk
gauge couplings and hopping superpotential are always turned on at the
stable fixed points.  We have checked this explicitly in
a large number of three-site models, and we have argued more generally 
in the braneless case that
the stable fixed points prefer the bulk gauge couplings and hopping 
superpotential to be turned on.  We expect this pattern to be completely
generic in the class of moose models described in this paper.
We conclude from these arguments that moose models with fundamental scalar
link fields {\em may} indeed provide 
a UV completion of an extra dimension.  
However, it does not follow that such an interpretation is 
possible without fine tuning.  We
are generally unable to calculate the values of the
nonvanishing couplings at the fixed point.  The NSVZ
$\beta$-functions depend on the anomalous dimensions, which
themselves depend on the gauge couplings.  At the fixed point, the 
$\beta$-functions vanish, which then determines the gauge couplings.
Near the edge of the conformal window where the fixed point
couplings are weak we showed that the gauge couplings 
are all equal at the stable fixed point in the
absence of a brane.
Although we do not know how to calculate the gauge couplings 
at strongly coupled fixed points, it is clear from the a-maximization analysis
that there is no dependence on the number of lattice
sites $k$ in the absence of a brane, and little dependence with a brane.  
Recall that for a deconstructed extra dimension with fixed 5D gauge coupling,
the 4D couplings should grow as $g_i\sim \sqrt{k}$ as $k$ is made large.  
Generally, then, we find that for fixed $k$ the effective higher dimensional 
gauge theory will be weakly coupled.
This conclusion assumes that $\Lambda_{QCD}$
is not too much larger than $\Lambda_{KK}$.  If $\Lambda_{QCD}\gg\Lambda_{KK}$
then it is possible for the theory to flow to still stronger coupling at
the scale $\Lambda_{KK}$.
However, if a hierarchy between dynamical scales must be imposed, then
the use of SUSY for preventing a mass hierarchy in the
deconstructed theory is made moot.  

Another challenge to the extra-dimensional interpretation is that
the superpotential couplings may still need to be
fine tuned to generate the correct hopping.  We have only demonstrated that
the superpotential is composed of relevant operators near the conformal
fixed points; we
have not calculated the superpotential
couplings in the IR.  To produce the spectrum of a
5D SUSY theory, the superpotential couplings may have to be tuned at $m_{KK}$.
It is interesting in this regard to note that, at least for the braneless
moose,
marginality of the trilinear superpotential couplings at the fixed points
automatically ensure that the mass terms for the flavors are also marginal.  
Hence, one might expect that those mass terms can be varied
freely along a continuum of fixed points \cite{leigh-strassler} 
and must therefore be tuned in order for the flavors to 
propagate at the same speed along the extra dimension as the gauge fields.  The
problem with this argument is that it does not take into account
the RG flows down to $m_{KK}$ when the  $\left<Q_i\right>$ VEVs are
turned on. 
Furthermore, the superpotential is arranged so that the low energy
theory below $m_{KK}$ has ${\cal N}$=2 supersymmetry in four dimensions, as 
opposed to the ${\cal N}$=1 supersymmetry of the high energy theory.  One may
hope that the true stable ground states after the $\left<Q_i\right>$ VEVs are
turned on would have this additional supersymmetry, although a conclusive
argument along these lines is lacking.  

The final point that we would like to make regarding RG flows and
deconstructed dimensions
is that some aspects of the
geometry and topology
of the deconstructed theory are determined dynamically in these theories.
The couplings which enter
the effective theory below the scale $\Lambda_{KK}$ are determined by their
values at the stable fixed point and the ratio $\Lambda_{QCD}/\Lambda_{KK}$.  
If additional dynamics also selects the link field vevs, then
the geometry of the deconstructed theory is completely determined.  
Furthermore, the boundary conditions, and hence the topology of the
extra dimension, are also determined dynamically.
We have found that whether or not the gauge coupling at the brane
lattice site vanishes at the stable
IR fixed point depends on the numbers of flavors and colors, 
at least in the cases with a small number of lattice sites which we have
studied in detail.  We expect this phenomenon to be generic in moose models
with conformal fixed points.  If the gauge coupling at
the brane is turned on at the stable fixed point, 
then the gauge fields have periodic boundary 
conditions on the
extra dimensional circle; if the gauge coupling is turned off, then the
gauge fields  effectively propagate on an interval.  Also,
in this case the brane-localized flavors decouple from the theory, so in 
essence the brane disappears.  The dynamics of the bulk flavors may
still need to be tuned in order to be consistent with the extra dimensional
interpretation and with the topology as seen by the gauge fields.  Indeed,
we have found that all of the flavor hopping terms are relevant operators
near the fixed points, so that the hopping terms at and 
near the brane are expected generically to be turned on.

There are several interesting
phenomena which may be the result of fixed point dynamics in moose models.
For example, in our analysis of three-site
models we have noticed that it 
tends to be models with {\em small} numbers of flavors for which the brane
gauge coupling remains turned on at the stable fixed point.  Hence,
flavor physics may be constrained in such models.

It might prove interesting to consider deconstruction of
5D product gauge group models
to explore the possibility of partial
breaking of the 5D gauge group via dynamically determined boundary conditions.
This would then provide a new dynamical realization of Higgsless type 
models \cite{Higgsless} or models of GUT breaking by
boundary conditions \cite{GUTs}.
Similar considerations may also lead to a dynamical realization of
partial breaking of 5D supersymmetry by the dynamically generated
boundary conditions, similar to Scherk-Schwarz SUSY breaking \cite{SS}.  
In particular, IR stability of the superpotential couplings
is not {\em a priori} linked to stability of the gauge couplings so that
the flavors might not be coupled to the brane in a manner consistent with
5D SUSY.

\section{Conclusions}
Moose models with conformal
fixed points have a rich
phenomenology which this paper only begins to unsurface.  
The discrete nature of
conformal fixed points makes them especially predictive, and in the
context of deconstructed extra dimensions can lead to a qualitative 
understanding of the theory at low energies.
For example, we demonstrated a dynamical selection of boundary conditions at 
the brane: the UV completion of the theory
dynamically selects the boundary conditions in the effective higher dimensional
theory, and thereby the topology of the extra dimension.

In addition to the rich phenomenology of deformed conformal moose models 
and their relation to extra dimensions, the 
intricate structure of conformal fixed points in these theories provides
a nontrivial class of tests of Cardy's conjectured
a-theorem.  In all cases studied, we have
found that the RG flows between fixed points are such that
the Euler anomaly $a$ flows from higher values to lower values as energy scale
decreases, in agreement
with Cardy's conjecture.  An analytic proof that $a$ always
flows from larger to smaller values in the context of flows between these
fixed points seems possible, although we have not provided such a proof
here.  It would be straightforward to generalize the study to
other product gauge group theories with conformal fixed points, 
  for example to theories with
additional ``branes'' or to the interval moose model
in its conformal window.

\section{Acknowledgments}This work
is supported in part by NSF Grant \# PHY-0504442 
and Jeffress Grant \# J-768. We thank
Tanmoy Bhattacharya, Christopher Carone, 
Ken Intriligator, Marc Sher, Yuri Shirman, and Neal Weiner for useful 
discussions.  JE acknowledges the Aspen Center for Physics, where some earlier
parts of this work were completed.


\end{document}